\documentclass[preprint, aps, pre, eqsecnum,amsmath,amssymb,showpacs]{revtex4}              

\usepackage[final]{graphicx}

\newcommand{\Nabla}{\mbox{\bf\boldmath $\nabla$}}

\newcommand{\bean}{\begin{eqnarray}}
\newcommand{\eean}{\end{eqnarray}}
\newcommand{\bea}{\begin{eqnarray*}}
\newcommand{\eea}{\end{eqnarray*}}
\newcommand{\beq}{\begin{equation}}
\newcommand{\eeq}{\end{equation}}

\def\lesssim{\mathrel{\mathpalette\vereq<}}
\def\vereq#1#2{\lower3pt\vbox{\baselineskip1.5pt \lineskip1.5pt
\ialign{$\hfill##\hfil$\crcr#2\crcr\sim\crcr}}}
\def\gtrsim{\mathrel{\mathpalette\vereq>}}

\begin{document}

\title{\bf Spiral Vortices and Taylor Vortices 
in the Annulus between Rotating Cylinders and the Effect of an Axial Flow}

\author{Ch.~Hoffmann, M.~L\"{u}cke, and A.~Pinter}
\affiliation{Institut f\"{u}r Theoretische Physik, \\ Universit\"{a}t
des Saarlandes, D-66041~Saarbr\"{u}cken, Germany\\}

\date{\today}

\begin{abstract}
We present numerical simulations of vortices that appear via primary
bifurcations out of the unstructured circular Couette flow in the Taylor-Couette
system with counter-rotating as well as with co-rotating cylinders. The full, 
time dependent 
Navier-Stokes equations are solved with a combination of a finite difference 
and a Galerkin method for a fixed axial periodicity length of the vortex 
patterns and for a finite system of aspect ratio 12 with rigid nonrotating ends 
in a setup with radius ratio $\eta=0.5$. Differences in structure,
dynamics, symmetry properties, bifurcation and stability behavior between
spiral vortices with azimuthal wave numbers $M=\pm1$ and $M=0$ Taylor vortices 
are elucidated and compared in quantitative detail. Simulations in axially
periodic systems and in finite systems with stationary rigid ends are compared
with experimental spiral data.
In a second part of the paper we determine how the above listed 
properties of the $M=-1,0,1$ 
vortex structures are changed by an externally imposed axial through-flow
with Reynolds numbers in the range $-40 \le Re \le 40$.
Among others we investigate when left handed or right handed spirals or
toroidally closed vortices are preferred.   
\end{abstract}


\maketitle
\clearpage

\section{Introduction}
Spiral vortices appearing in the annular gap between the concentric rotating
cylinders of the Taylor-Couette system \cite{T94} are a rather interesting 
example for the
spontaneous formation of a helicoidal structure out of a homogeneous state of a 
nonlinear forced
system when the forcing exceeds a critical threshold. Like the competing
toroidally closed Taylor vortices the spiral vortex structures
bifurcate out of the unstructured basic state of circular Couette flow (CCF) 
that is
stable at small rotation rates of the inner cylinder. The spiral pattern breaks
the rotational symmetry of the annular gap. It oscillates in time by rotating
azimuthally as a whole thereby propagating axially. The Taylor vortex flow
(TVF), on the other hand is rotationally symmetric and stationary. 

\begin{figure}[ht]
\begin{center}
\includegraphics[width=8.5cm]{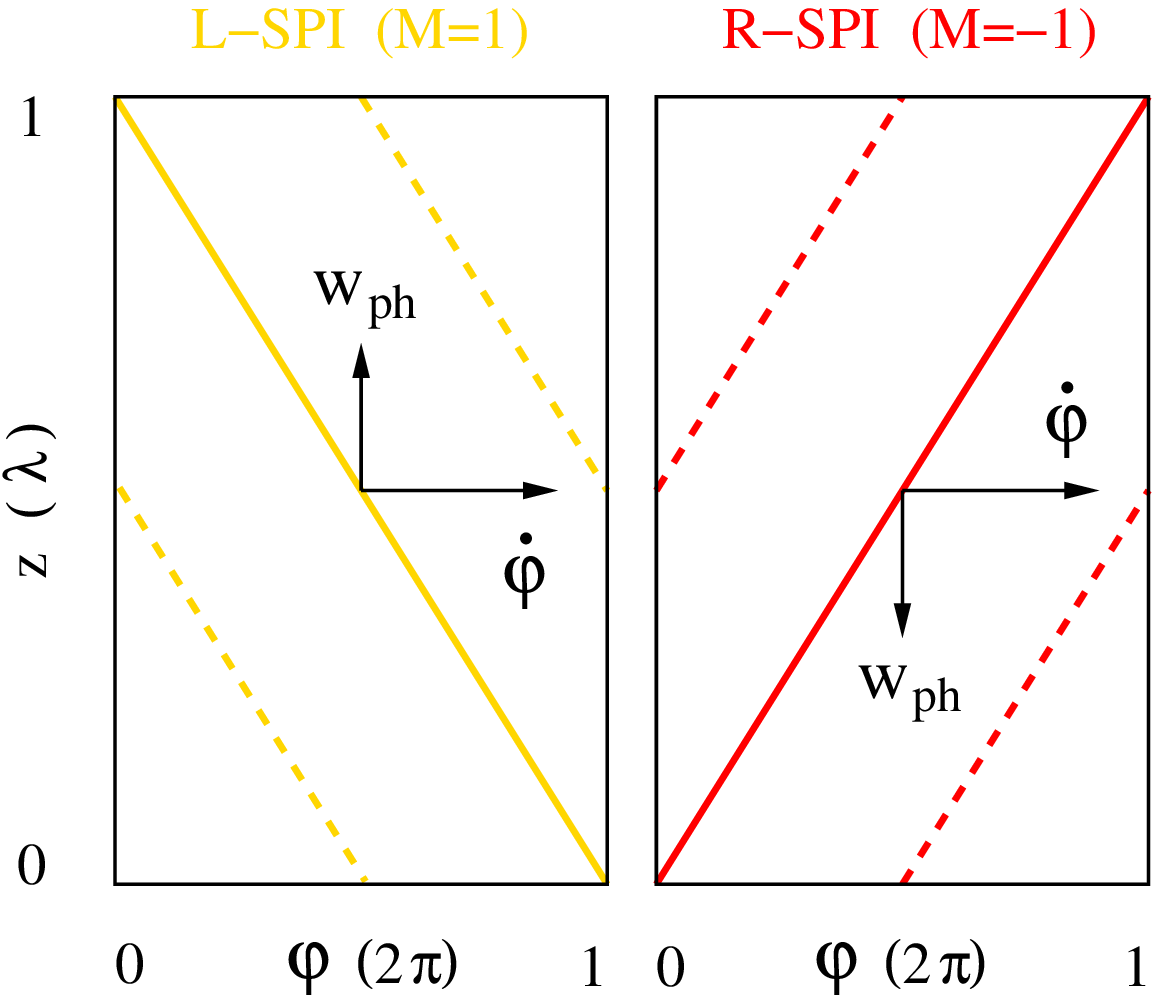}
\end{center}
\caption{Lines of constant phases, $\phi=const$, for spirals in the 
$\varphi - z$ plane. Arrows indicate their velocities.}
\label{FIG:phases}
\end{figure}

The spiral pattern is effectively one dimensional (like TVF) and stationary
when seen from a co-moving frame \cite{CI94}: the spiral fields do not depend on 
time $t$,
axial coordinate $z$, and azimuthal angle $\varphi$ separately but only via the
combined phase variable $\phi= kz + M\varphi -\omega(k, M) t$. Here $k$ and $M$
are the axial and azimuthal wave numbers, respectively, and $\omega$ the
frequency. In the $\varphi-z$ plane of an 'unrolled' cylindrical surface the
lines of constant phase, $\phi = \phi_0$, are straight with slope $-M/k$ as
shown in Fig.~\ref{FIG:phases}. 
An azimuthal wave
number $M>0$ implies a left handed spiral (L-SPI) while $M<0$ refer to
right handed spirals (R-SPI) with our convention of taking $k$ to be positive.
L-SPI and R-SPI being mirror images of each
other under the operation $z \to -z$ are symmetry degenerate flow states. Which
of them is realized in a particular experimental or numerical setup depends on
the initial conditions.

With the lines of constant phase in the $\varphi-z$ plane  being oriented for 
both spiral types obliquely to the azimuthal 'wind' of the basic CCF both
spirals are advectively rotated by the latter like rigid objects. Their common 
angular velocity is $\dot{\varphi}_{SPI} = \omega(k,M)/M$. 
This advection enforced rigid-body rotation of the spiral vortices is also
reflected by the fact that the axial phase velocities
$w_{ph} = \omega /k = \dot{\varphi}_{SPI} M /k$ of an L-SPI 
($M>0$) and of an R-SPI ($M<0$) are opposite to each other, see 
Fig.~\ref{FIG:phases}. 
By the same token the rotationally symmetric ($M=0$) structure of toroidally 
closed Taylor vortices is stationary ($\omega=0$): 
the lines of constant phases being parallel to the azimuthal CCF the latter
cannot advect these vortices. However, an externally imposed axial 
through-flow can advect Taylor vortices as well as spiral vortices.

The external through-flow breaks the mirror symmetry between L-SPI and R-SPI.
It changes their rotation and propagation dynamics as well as their structural
properties and their bifurcation behavior in different ways. This is the topic
of our investigation.  

In his review \cite{T94} Tagg remarks that systematic investigation of
non-axisymmetric vortex states that appear via primary bifurcations out of the CCF
state started remarkably late in the history of the Taylor-Couette problem.
Krueger et al. \cite{KGD66} predicted in 1966 primary transitions to 
non-axisymmetric
rotating-wave flow which then were observed in experiments by Snyder
\cite{S68} who had presented experimental evidence for different types of
stable helical flow (referred to as 'spirals') a few years earlier. In 1985,
an experimental survey was published by Andereck et al. \cite{ALS85}
which classified a large variety of different flow states, including some
spiral types like linear, modulated, interpenetrating, and wavy spirals etc.
An extensive numerical linear stability analysis was then performed for a
wide range of radius ratios by Langford et al. \cite{LTKSG88}. At this
time, Tagg et al. \cite{TESM88} experimentally observed a transition from
CCF to axially standing and azimuthally
traveling waves (ribbons) and found numerically calculated wave speeds to be
in agreement with experimental results. Edwards \cite{E90} studied the
transition from CCF to traveling waves. More recent experiments were done 
with a system of radius ratio $\eta=0.5$ \cite{SP99}. Spiral solutions in a 
relatively narrow gap
with axially periodic boundary conditions were obtained numerically with a 
pseudo-spectral method using co-rotating helicoidal coordinates which were
adapted to the expected spiral \cite{AMS98}.

Various effects of an externally imposed axial through-flow in the Taylor-Couette
system have been explored since the early 1930 so that the list of publications
cannot be discussed here -- see, e.g, Ref.~\cite{BLRS96} for a 
partial and far from complete compilation. We mention here in addition
a few, more recent papers on selected topics beyond those listed in 
Ref.~\cite{BLRS96}:
linear analysis of the competition between shear and centrifugal instability
mechanisms \cite{GG93,MM02}; linear SPI and TVF fronts and
pulses \cite{PLH03}; weakly nonlinear bifurcation analysis of axially extended
spiral, ribbon, and mixed vortex states with homogeneous amplitudes
\cite{RL93,CI94}; theoretical/numerical investigation of the nonlinear pattern
selection in the absolutely unstable regime under downstream evolving intensity
envelopes \cite{BLRS96}; theoretical/numerical analysis of noise-sustained
patterns in the convectively unstable regime \cite{noise-effects} (related 
experiments are listed in \cite{BLRS96}); analysis of the changes
in the noise sensitivity across the convective-absolute stability boundary 
\cite{noise-sensitivity};  measurements of velocity fields by particle image
velocimetry \cite{WL99}.

In this work we explore in a detailed quantitative investigation the
spatio-temporal structures as well as the bifurcation properties of spirals and
TVF in a setup with co- and
counter-rotating cylinders of fixed radius ratio $\eta=0.5$ with and without an
externally imposed axial through-flow. Most calculations were done for
axially periodic boundary conditions that impose the wave length of the vortex
pattern. However also a few simulations of finite systems with rigid, 
non-rotating lids were done to compare with experiments and to study the effect
of phase propagation suppressing boundaries.  
The calculations were done with a time dependent finite differences 
method in the $r-z$ plane combined with a spectral decomposition in $\varphi$
which yields by construction only the stable flows. However, by selectively 
suppressing destabilizing modes we also were able to trace out the unstable 
TVF and SPI solution branches. We do not include in this work results on 
ribbons \cite{TESM88}, i.e., nonlinear combinations of L and R spirals 
\cite{CI94} since they were unstable for the parameters investigated here.

In Sec.~\ref{SEC:System} we present the notation for describing the Taylor
Couette system and we describe our numerical method. In Sec.~\ref{SEC:SPI-TVF}
we review the spatio-temporal properties of TVF and SPI and we present results
on their bifurcation behavior and flow structure in the absence of 
through-flow. In particular we provide detailed comparisons of the bifurcation
and structural properties of these primary vortex states. Also comparisons with
experiments are presented and discussed. 
In Sec.~\ref{SEC:ETF} we elucidate the effect of an external through-flow on
structure, dynamics, and bifurcation properties of TVF and SPI for counter-rotating
cylinders and stationary outer cylinder. The last section contains a summary of
the main results. 

\section{System} \label{SEC:System}
We report results obtained numerically for a Taylor-Couette system with co-
and counter-rotating cylinders. The ratio $\eta=r_1/r_2$ of the radii $r_1$
and $r_2$ of the inner and outer cylinders, respectively, was fixed at the 
value $\eta=0.5$ for which also experiments have been made recently 
\cite{SP99}. 

\subsection{Theoretical description}
We consider the fluid in the annulus between the cylinders to be isothermal
and incompressible with kinematic viscosity $\nu$. The gap width $d=r_2-r_1$
is used as the unit of length and the momentum diffusion time $d^2/ \nu$
radially across the gap as the time unit so that velocities are reduced by
$\nu /d$. To characterize the driving of the system, we use the Reynolds
numbers
\begin{eqnarray}
R_1=r_1\Omega_1 d/\nu \,\, ; R_2=r_2\Omega_2 d/\nu \, . 
\end{eqnarray}
They are just the reduced azimuthal velocities of the fluid at the inner and
outer cylinder, respectively, where $\Omega_1$ and $\Omega_2$ are the
respective angular velocities of the cylinders. The inner one is always
rotating counterclockwise so that $\Omega_1$ and $R_1$ are positive. 
We explore positive as well as negative values of $R_2$ 
corresponding to co- as well as counter rotation of the cylinders, 
respectively. And we elucidate the effect of an externally imposed axial 
through-flow. 

Within the above described scaling, the NSE take the form
\begin{eqnarray}
\partial_t {\bf u} = \Nabla^2 {\bf u} -
({\bf u}\cdot \Nabla){\bf u} - \Nabla p \,. 
\end{eqnarray} 
Here $p$ denotes the pressure reduced by $\rho \nu^2/d^2$ and $\rho$ is the mass 
density of the fluid. Using cylindrical coordinates, the velocity field
\begin{eqnarray}
{\bf u}=u\,{\bf e}_r + v\,{\bf e}_\varphi + w\,{\bf e}_z 
\end{eqnarray}
is decomposed into a radial component $u$, an azimuthal one $v$, and an
axial one $w$. 

We have solved the resulting equations subject to no slip conditions at the
cylinders. In Sec.~\ref{SEC:CER} we
present simulations of systems with axial size $\Gamma=12$ and
rigid stationary ends bounding the annulus axially in order to compare
with experiments \cite{SP99}. For the main part (c.f. Secs.~\ref{SEC:SPI-TVF}
and \ref{SEC:ETF}) of this work we imposed, however, axially periodic boundary
conditions at $z=0$ and $z=\Gamma=1.6$. So the axial wavelength of the TVF
and the SPI patterns investigated there is $\lambda=1.6$ and the  wave number
is $k=2\pi/\lambda=3.927$.

\subsection{Numerical method}\label{SEC:Num-method}
The calculations were done with a finite differences 
method in the $r-z$ plane combined with a spectral decomposition in $\varphi$.
Since we have been studying also finite length cylinders, say, with lids
bounding  
the annulus vertically, we do not use here a spectral decomposition in axial 
direction which for axially periodic systems is a generic alternative.
The discretization (a FTCS - Forward Time, Centered
Space algorithm) has been done on staggered grids in the $r-z$ plane
following the procedure of Ref. \cite{HNR75}. It yields simple
expressions for the derivatives, it does not require boundary conditions for the 
pressure, and it avoids difficulties with boundary conditions for more than one 
velocity field component at the same position.
We used homogeneous grids with discretization lengths $\Delta r=\Delta
z=0.05$ which have showed to be more accurate than non-homogeneous grids.
Time steps were $\Delta t<1/3600$. 

Azimuthally all fields $f=u,v,w,p$ were expanded as
\begin{eqnarray}\label{EQ-expansion}
f(r,\varphi,z,t)=
\sum_{m=-m_{max}}^{m_{max}} f_m(r,z,t)\,e^{im\varphi} .
\end{eqnarray}
For the flows investigated here a truncation of the above Fourier expansion
at $m_{max}=8$ was sufficient to properly resolve the anharmonicities in the
fields. The system of coupled equations for
the amplitudes $f_m(r,z,t)$ of the azimuthal normal modes
$-m_{max} \leq m \leq m_{max}$ is solved with the FTCS algorithm.
Pressure and velocity fields are iteratively adjusted to each other with the 
method of 'artificial compressibility' \cite{PT83}
\begin{eqnarray}
\label{dp-n} 
dp^{(n)}        &=& - \beta \, \Nabla \cdot {\bf u}^{(n)} \qquad (0<\beta<1) \\
p^{(n+1)}       &=& p^{(n)} + dp^{(n)} \\
\label{u-nplus1}
{\bf u}^{(n+1)} &=& {\bf u}^{(n)} - \Delta t \, \Nabla (dp^{(n)}) \,.
\end{eqnarray}
The pressure correction $dp^{(n)}$ in the $n$-th
iteration step being proportional to the divergence of ${\bf u}^{(n)}$
is used to adapt the velocity field ${\bf u}^{(n+1)}$.
The iteration loop (\ref{dp-n}-\ref{u-nplus1}) is executed for each azimuthal 
Fourier mode separately. It is iterated until $\Nabla \cdot {\bf u}$ has 
become sufficiently small for each $m$ mode considered -- 
the magnitude of the total divergence never exceeded 0.02 and typically 
it was much smaller. After that the next FTCS time step was executed.

For code validation we compared SPI solutions with experiments \cite{SP99} and 
TVF solutions with previous numerical simulations \cite{BLRS96} and close to
onset also with Ginzburg-Landau results \cite{REC-LUE-MUE}. Furthermore, we
compared bifurcation thresholds of the nonlinear SPI and TVF solutions with
the respective stability boundaries of the linearized NSE \cite{LTKSG88,PLH03} 
obtained by a shooting method that is described in detail in 
\cite{PLH03}. As expected from our experience with primary vortex
structures in the Taylor-Couette and Rayleigh-Benard problem lie the MAC FTCS
bifurcation thresholds for our discretization typically 1 - 2 \% below the
respective linear stability thresholds. This deviation significantly reduces 
for finer discretizations. We also investigated how the nonlinear solutions
change when varying $m_{max}$ and/or the grid spacing. From these analyses we
conservatively conclude that typical SPI frequencies have an error of less than
about 0.2\% and that typical velocity field amplitudes can be off by about
3 - 4\%. Time steps were always well below the von Neumann stability
criterion and by more than a factor of three below the Courant-Friederichs-Lewy
criterion.

In order to trace out the unstable parts of bifurcation branches of TVF and SPI
solutions we applied different stabilization methods that are described in 
Sec.~\ref{RFA}.   

\section{Spiral vortices and Taylor vortices}
\label{SEC:SPI-TVF}
In this section we first briefly review spatio-temporal properties of 
spiral vortices ($M\neq0$) and Taylor vortices ($M=0$) in the absence of any
externally enforced axial through-flow. Here $M$ is the azimuthal wave number
of the respective vortex structure. 
Then we present our results on the bifurcation behavior of $M=0$ and $M=\pm1$
vortex solutions and on their flow structure.  

They both grow out of the basic CCF state, 
${\bf u}_{CCF} = v_{CCF}(r){\bf e}_{\varphi}$, that is rotationally symmetric,
axially homogeneous, and time translationally invariant. Here in our system 
with $\eta=1/2$ the radial profile of its azimuthal velocity reads
\begin{equation}\label{v_CCF}
v_{CCF}(r) =\frac{2R_2-R_1}{3}r + \frac{4R_1-2R_2}{3}\frac{1}{r} \, .
\end{equation} 

\subsection{Spatio-temporal structure}

The spiral vortex structure is periodic in $\varphi,z$, and $t$. It
rotates uniformly as a whole like a rigid object in 
azimuthal direction thereby translating with constant phase velocity in axial 
direction --- the spiral fields $f(r,\varphi,z,t)$ do not depend on 
$\varphi,z,t$ separately but only the phase combination
\begin{equation}\label{phase}
\phi= kz + M\varphi -\omega(k, M) t \, .
\end{equation}
Here $k$ is the axial wave number that we always take to be positive and 
$\omega(k, M)$ is the frequency. Thus, with 
$f(r,\varphi,z,t) = F(r, \phi)$, the spiral pattern is 
one dimensional. Comparing the Fourier decompositions 
\begin{subequations}
\begin{eqnarray}
f(r,\varphi,z,t) = \sum_{m,n} f_{m,n}(r,t)\,e^{i(m\varphi + nkz)} 
= \sum_{\nu} F_{\nu}(r)\,e^{i\nu \phi} = F(r,\phi)
\end{eqnarray}
one finds that
\begin{eqnarray}
f_{m,n}(r,t) = \delta_{m,nM} \, e^{-i n\omega t} F_n(r)\,. 
\end{eqnarray}
\end{subequations}
Thus only the mode combinations $m=nM$ appear in a SPI with azimuthal wave
number $M$.

The SPI phase is constant, $\phi_0$, on a cylindrical surface,
$r=const$, along lines given by the equation
\begin{equation} \label{z_phase}
z_0 = - \frac{M}{k}\varphi + \frac{\omega(k,M)}{k}t + \frac{1}{k}\phi_0 \, .
\end{equation}
Thus, on the $\varphi-z$ plane of such an 'unrolled' cylindrical surface these
lines of constant phase are straight with slope $-M/k$. And an azimuthal wave
number $M>0$ implies a left handed spiral (L-SPI) while $M<0$ refer to
right handed spirals (R-SPI) with our convention of taking $k$ to be positive.
L-SPI and R-SPI being mirror images of each
other under the operation $z \to -z$ are symmetry degenerate flow states. Which
of them is realized in a particular experimental or numerical setup depends on
the initial conditions.

The lines of constant phase and with it the whole spiral structure rotates 
in $\varphi$ with angular velocity
\begin{equation} \label{phidot_phase}
\dot{\varphi}_{SPI} = \frac{\omega}{M} \, .
\end{equation}
Its direction strongly depends on the inner cylinder's rotation due to the 
influence of the CCF. The latter decisively determines the shape of the  
linear spiral eigenmodes that can grow beyond the stability boundary of the CCF state
against perturbations with azimuthal wave number $M\neq0$. In the
parameter range explored here the spirals rotate into the same direction as the
inner cylinder, i.e., into positive $\varphi$-direction so that 
$\omega(k, M)/M$ is always positive, i.e., $\omega = sign (M) |\omega|$.
From this rigid rotation one immediately infers from Eq.(\ref{phase}) that
the axial phase velocity
\begin{equation} 
w_{ph} = \frac{\omega}{k} = \frac{M}{k} \dot{\varphi}_{SPI}
\end{equation}
of an L-SPI ($M>0$) is positive and of an R-SPI ($M<0$) it is negative. 

For the rotationally symmetric ($M=0$) structure of toroidally closed Taylor 
vortices the lines of constant phases are parallel to ${\bf e}_\varphi$. 
This $M=0$ pattern is stationary ($\omega=0$). The main reason is that
the azimuthal flow of the basic CCF state being precisely parallel to the 
vortex lines of constant phase cannot advect them. However, an  axial 
mean flow, being perpendicular to them can advect them: an externally
enforced axial through-flow of strength $Re$ causes a non-zero 
axial phase velocity of the Taylor vortex pattern that grows linearly with $Re$,
at least when phase pinning
effects are absent as for axially periodic boundary conditions.

\subsection{Bifurcation behavior}
In the parameter regime considered here the bifurcation thresholds for
nonlinear SPI and TVF solutions, i.e., the linear stability boundaries of the 
CCF state against 
$M=\pm 1$ and $M=0$ vortex perturbations \cite{LTKSG88} differ only slightly 
from each other. For our fixed wave number of $k=3.927$ they intersect at 
$(R_1^s=95.25, R_2^s=-73.69)$ where these two different vortex modes are
bi-"critical" in the sense that their growth rates are simultaneously zero.
The stability boundaries were obtained with a shooting method from the
linearized NSE. The nonlinear SPI and TVF solutions that were determined
with the numerical method described in Sec.~\ref{SEC:Num-method} give
bifurcation thresholds that differ as a result of the FTCS discretization
errors by at most 2\% from the linear stability analysis. However, this 
difference can grow with externally applied through-flow up to, say, 5\% 
at $Re\simeq40$ (c.f. Sec.~\ref{SEC:ETF}) when the discretization is not refined.

\subsubsection{Radial flow amplitudes of TVF and SPI}
\label{RFA}

\begin{figure}[ht]
\begin{center}
\includegraphics[width=14.5cm]{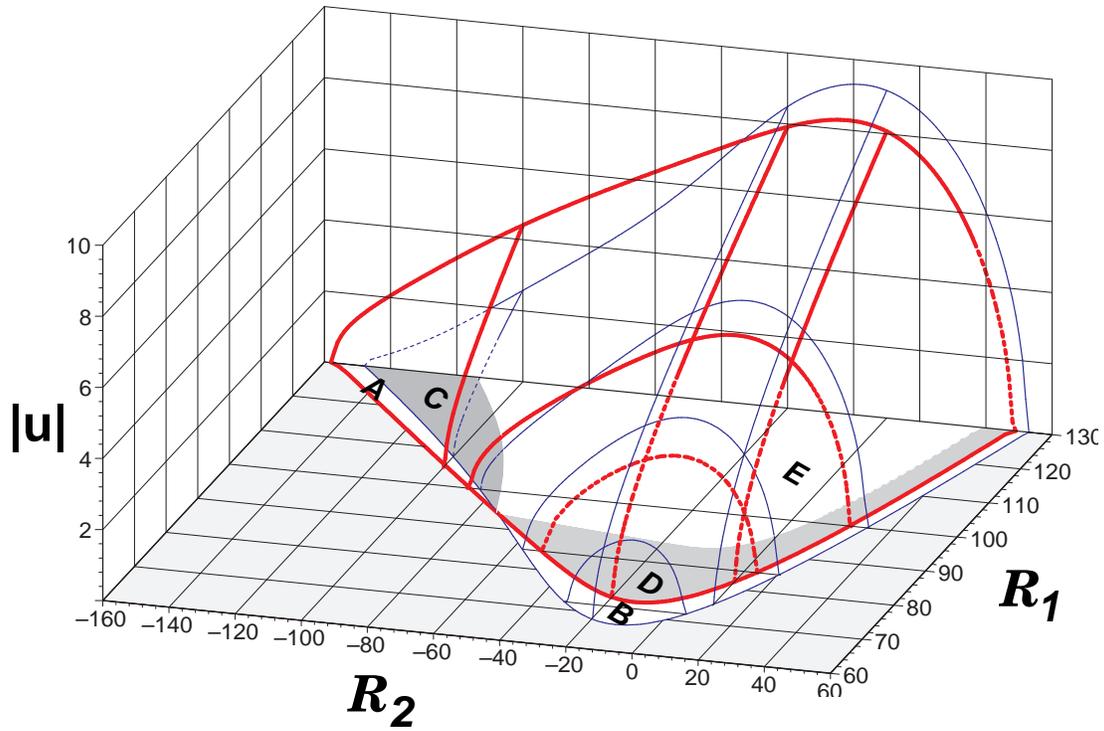}
\end{center}
\caption{ 
Order parameter bifurcation surfaces of TVF (thin lines) and of 
SPI (thick lines) over the $R_1 - R_2$ plane. Shown are primary 
Fourier amplitudes, $|u_{m,n}|$, of the radial flow intensity at mid gap, 
$r=r_1 + 0.5$, with axial mode index $n=\pm 1$. The azimuthal one is $m=0$ for
TVF and $m =\pm 1$ for SPI, respectively. In each case full (dashed) lines 
denote stable (unstable) solutions.} 
\begin{tabular}{|c|c|c|c|c|c|}
\hline
region & A      & B      & C        & D        & E      \\\hline\hline
TVF    & -      & stable & unstable & stable   & stable \\\hline
SPI    & stable & -      & stable   & unstable & stable \\\hline
\end{tabular}
\label{FIG:BIF-u}
\end{figure}

The bifurcation of both, TVF and SPI solutions is forward as
shown by the bifurcation surface over the $R_1 - R_2$ plane of Fig.~\ref{FIG:BIF-u}. 
There the respective vortex solution is characterized by the primary 
Fourier amplitude, $|u_{m,n}|$, of the radial flow intensity at mid gap, 
$r=r_1 + 0.5$, taken
as order parameter with $m$ denoting the azimuthal mode index and $n$ referring
to the axial one, respectively. Thus, Fig.~\ref{FIG:BIF-u} shows $|u_{0,1}|$ for 
the TVF solution by thin lines and $|u_{1,1}|=|u_{-1,1}|$ for the two symmetry 
degenerate $M=\pm 1$ SPI solutions by thick lines, respectively. In each
case stable (unstable) solutions are represented by full (dashed) lines. 
The different stability regions labelled A - E are explained in the caption of
Fig.~\ref{FIG:BIF-u}.

The stability of the vortex states refers to our system with fixed axial
periodicity length. Thus, e.g., Eckhaus or Benjamin-Feir instabilities 
\cite{CRO-HOH} that can destabilize periodic patterns in infinite and large 
systems do not occur here. Furthermore, our periodic
boundary conditions allowing free phase propagation enhance the
existence range as well as the stability range of SPI solutions in comparison
with, say, Ekman vortex generating stationary lids that axially close the 
annulus in an experimental setup. The latter suppress phase propagation 
in their vicinity so that phase generating and phase destroying defects near
opposite boundaries are necessary for the realization of spirals in the bulk 
of such systems.

In our setup TVF is for $R_2 > R_2^{s}$ stable close to onset. And it remains 
so at least up to the
largest value of $R_1=130$ shown in Fig.~\ref{FIG:BIF-u} -- for larger $R_1$ TVF
eventually undergoes an oscillatory instability. For more negative 
$R_2 < R_2^{s}$ 
TVF is unstable at onset (region C in Fig.~\ref{FIG:BIF-u}) but becomes stable at 
larger $R_1$ in region E. The unstable TVF solution branch in region C was 
obtained by suppressing any $m\neq0$ modes in the field
representation (\ref{EQ-expansion}), i.e., by allowing only rotationally
symmetric solutions. Lifting this mode restriction infinitesimal $m\neq0$
perturbations drive the system in the parameter region C of Fig.~\ref{FIG:BIF-u} 
away from the unstable TVF solution into a stable SPI state.

Spirals, on the other hand, are for $R_2 < R_2^{s}$ stable close to onset and 
remain so at least up to the largest value of $R_1=130$ shown in 
Fig.~\ref{FIG:BIF-u} while for $R_2 > R_2^{s}$ they are unstable at onset 
(region D in Fig.~\ref{FIG:BIF-u}). But then they become stable at larger  
$R_1$ in region E. The unstable SPI solution branch in region D was 
obtained by suppressing $m=0$ contributions to the radial velocity field $u$
at mid gap location. This stabilized the SPI solution against the growth of
TVF. Lifting this restriction of the available mode space the unstable SPI
solutions in region D decay into stable Taylor vortices.

In the relatively large region E both, SPI as well as TVF solutions coexist 
bistably and the final vortex structure to be found here depends on the initial 
conditions and the driving history of $R_1, R_2$. Note in particular that for
our periodic boundary conditions the region E with stable spirals extends to
positive $R_2$, i.e., to a situation with co-rotating cylinders.
 
\subsubsection{SPI frequencies}

\begin{figure}[ht]
\begin{center}
\includegraphics[width=14.5cm]{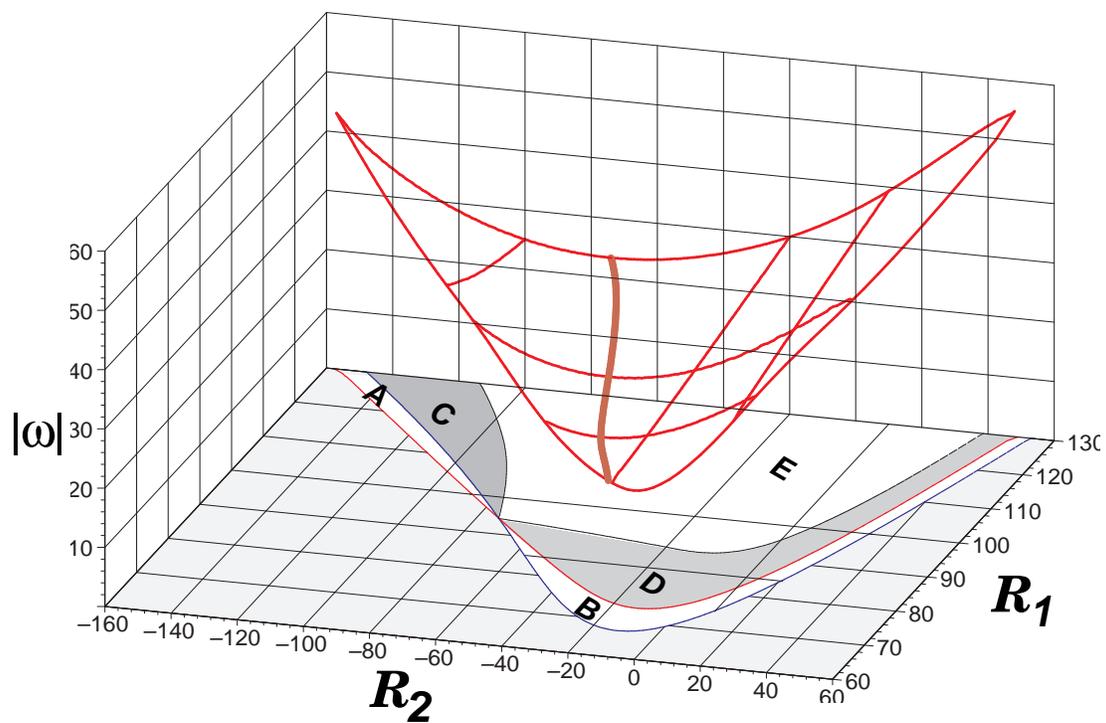} 
\end{center}
\begin{quote}
\caption{Bifurcation diagram of $M= \pm 1$ spiral frequencies $\omega$
over the $R_1 - R_2$ plane. The thick line locates the minima. 
The different stability regions A -E of TVF and 
of SPI solutions (c.f. caption of Fig.~\ref{FIG:BIF-u}) in the $R_1 - R_2$ plane 
are included for better comparison with Fig.~\ref{FIG:BIF-u}. }
\label{FIG:BIF-om}
\end{quote}  
\end{figure}

In Fig.~\ref{FIG:BIF-om} the spiral frequencies $\omega$ are plotted over the same 
control
parameter range as the radial flow amplitudes in Fig.~\ref{FIG:BIF-u}. Also
here we include -- for the sake of comparison with Fig.~\ref{FIG:BIF-u} -- the
identification of the different stability regions of TVF and SPI solution by 
the symbols A-E explained in the caption of Fig.~\ref{FIG:BIF-u}. At onset $\omega$
agrees within the numerical accuracy of our nonlinear code with the eigenvalue 
resulting from the linear stability analysis of the CCF state.

The nonlinear SPI frequencies further away from onset
vary smoothly: the bifurcation surface of $\omega$ in Fig.~\ref{FIG:BIF-om}
has the shape of a cloth that hangs down from a frame given by the linear 
onset spiral frequencies $\omega(R_{1,stab})$ at the stability threshold 
$R_{1,stab}(R_2)$ of CCF. The location of minimal $\omega$ on the 
bifurcation surface is shown by a thick line in Fig.~\ref{FIG:BIF-om}. 
Thus, the nonlinear SPI frequencies are typically smaller than the linear 
ones but do not deviate substantially from them. 

\begin{figure}[ht]
\begin{center}
\includegraphics[width=14.5cm]{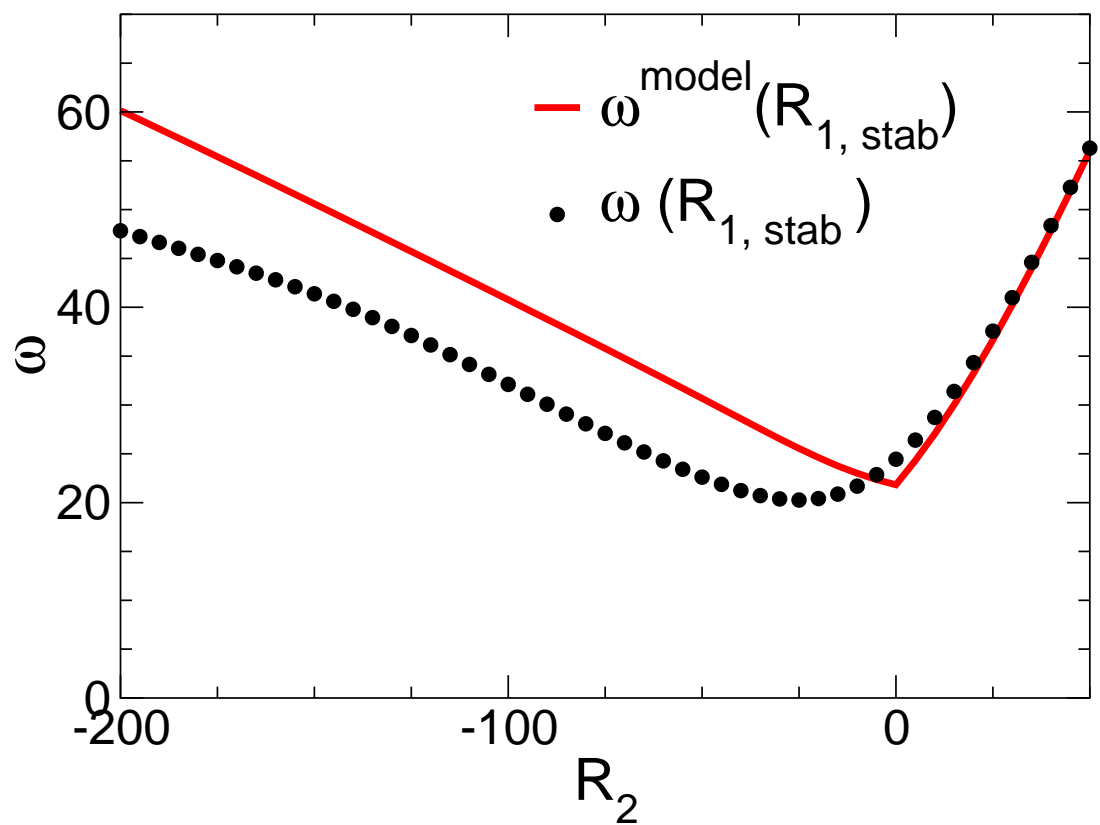}    
\end{center}
\begin{quote}
\caption{Linear frequency $\omega(R_{1,stab})$ of $M= 1$ spiral at onset, 
$R_{1,stab}(R_2)$, in comparison with frequency 
$\omega^{model}(R_{1,stab})$ (\ref{EQ:om-model}) resulting from rigid-body
rotation model.}
\label{FIG:om-model}
\end{quote}  
\end{figure}

Since the linear onset frequencies show a characteristic variation along the 
bifurcation threshold, $R_{1,stab}(R_2)$, that dictates the form of the
whole $\omega$ bifurcation surface we discuss them in some detail. They,
furthermore allow for a simple, yet semiquantitive 
explanation of the phenomenon of rigid body rotation of 
spirals in terms of a passive advection dynamics of $M = \pm 1$ vortex
perturbations, $e^{i\phi}$, with lines of constant phase, $\phi=kz +M\varphi
-\omega t$, that are oriented obliquely to the "wind" of the basic azimuthal 
CCF. To that end we compare in Fig.~\ref{FIG:om-model} the onset spiral
 frequency $\omega(R_{1,stab})$ at the stability threshold $R_{1,stab}(R_2)$
of CCF with the "model" frequency 
$\omega^{\mbox{model}}(R_{1,stab})$ which is 
evaluated also at the stability threshold $R_{1,stab}(R_2)$. Here 
\begin{eqnarray} \label{EQ:om-model}
\omega^{\mbox{model}}= \langle \omega_{CCF}(r) \rangle =
\frac{2}{r_0^2-r_1^2} \int\limits_{r_1}^{r_0}\omega_{CCF}(r)\,r\,dr
\end{eqnarray}
is the mean of the rotation rate of the CCF, $\omega_{CCF}=v_{CCF}/r$. For 
$R_2 < 0$ the averaging is done over
the radial domain between inner cylinder, $r_1$, and the first zero, $r_0$, 
of $v_{CCF}(r)$ (\ref{v_CCF}). Thus, at the stability threshold 
$R_{1,stab}(R_2)$ one has
\begin{eqnarray}
r_0^2= \frac{2R_2 - 4R_{1,stab}}{2R_2 - R_{1,stab}} 
\end{eqnarray}
when $R_2 < 0$. However, when $R_2 \geq 0$, i.e., when $v_{CCF}$ remains 
positive throughout the gap $r_0$ is replaced by $r_2$.
The restriction of the radial average to the range between $r_1$ and $r_0$ is
motivated by an argument of largely hand-waving nature:
the linear eigenfunctions for marginally stable SPI modes are
somewhat centered to this range where the growth of vortex perturbations is
supported. 

Fig.~\ref{FIG:om-model} shows that the onset spiral
frequency $\omega(R_{1,stab})$ agrees perfectly well with 
the mean CCF rotation frequency (\ref{EQ:om-model}) when $R_2 > 0$. For
$R_2 < 0$ the model ansatz (\ref{EQ:om-model}) for the global spiral rotation 
rate overestimates slightly the spiral 
frequency since Eq.(\ref{EQ:om-model}) does not contain contributions from 
negative CCF rotation rates between $r_0$ and $r_2$. In fact, if one extends 
in an ad hoc way the averaging domain slightly beyond $r_0$ then the agreement
improves significantly. Thus, the onset spiral frequency $\omega(R_{1,stab})$,
i.e., the frequency eigenvalue
can be seen as the mean rotation rate of the CCF -- albeit weighted 
appropriately by the critical eigenfunctions.  

\subsection{Flow structure of TVF and SPI}

\begin{figure}[ht]
\begin{center}
\includegraphics[width=12.5cm]{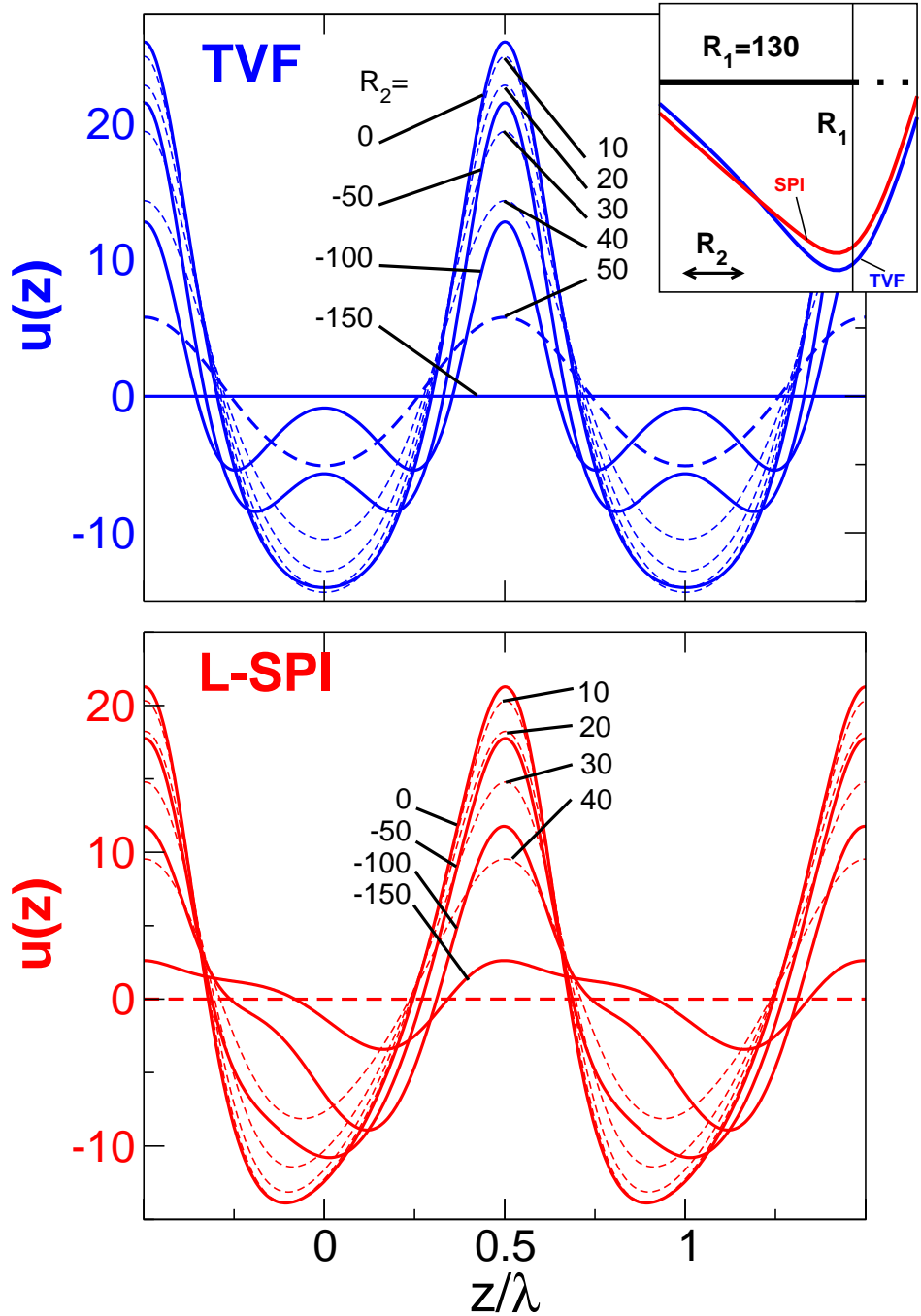}
\end{center}
\begin{quote}
\caption{
Axial profiles of the radial velocity $u(z)$ at mid gap position for $R_1 = 130$ and
various $R_2$ (along the thick horizontal line in the inset) covering the whole 
interval between the bifurcation thresholds marked TVF and SPI, respectively, in the
inset; see also Fig.~\ref{FIG:BIF-u}. Full (dashed) lines refer to negative
(positive) $R_2$. In each case the maximal radial outflow is 
chosen to lie at $z=0.5 \lambda$. For better visibility two axial periods of the 
vortex profiles are shown. The $M=1$ L-SPI are propagating in positive $z$-direction.
Parameters are $\eta=0.5$, $k=3.927$.}
\label{FIG:u-profiles}
\end{quote}  
\end{figure}

In this section we elucidate the flow structure of spiral vortices in comparison
with Taylor vortices. To that end we consider the radial velocity field. 
In Fig.~\ref{FIG:u-profiles} we show the axial profiles of 
$u(z)$ at mid gap position for $R_1 = 130$ being fixed and various $R_2$ that cover
the whole interval between the bifurcation thresholds, c.f. Fig.~\ref{FIG:BIF-u} 
and the inset of Fig.~\ref{FIG:u-profiles}. Full (dashed) lines refer to negative
(positive) $R_2$. In each case the axial position of maximal radial outflow is 
chosen to lie at $z=0.5 \lambda$. For the sake of better visibility two axial 
periods of the vortex profiles are shown.

\subsubsection{Anharmonicity: TVF versus SPI}

Typically SPI are less anharmonic than TVF. Also the profiles of both are
less anharmonic for positive $R_2$ than for negative $R_2$ and the degree of
anharmonicity increases when $R_2$ becomes more negative. For the mirror
symmetric TVF this anharmonicity growth comes from a widening (narrowing) of
the axial range $\Delta_{in}$ ($\Delta_{out}$) of radial inflow over which
$u<0$ ($u>0$) and the corresponding decrease (increase) of the inflow
(outflow) velocity. For the L-SPI that propagate in
Fig.~\ref{FIG:u-profiles} into positive $z$-direction the anharmonicity
grows mainly by flattening (steepening) of the wave profiles ahead of
(behind) the crests. However, $\Delta_{in} /
\Delta_{out}$ increases also for SPI albeit less than for TVF.

\begin{figure}[ht]
\begin{center}
\includegraphics[width=13.5cm]{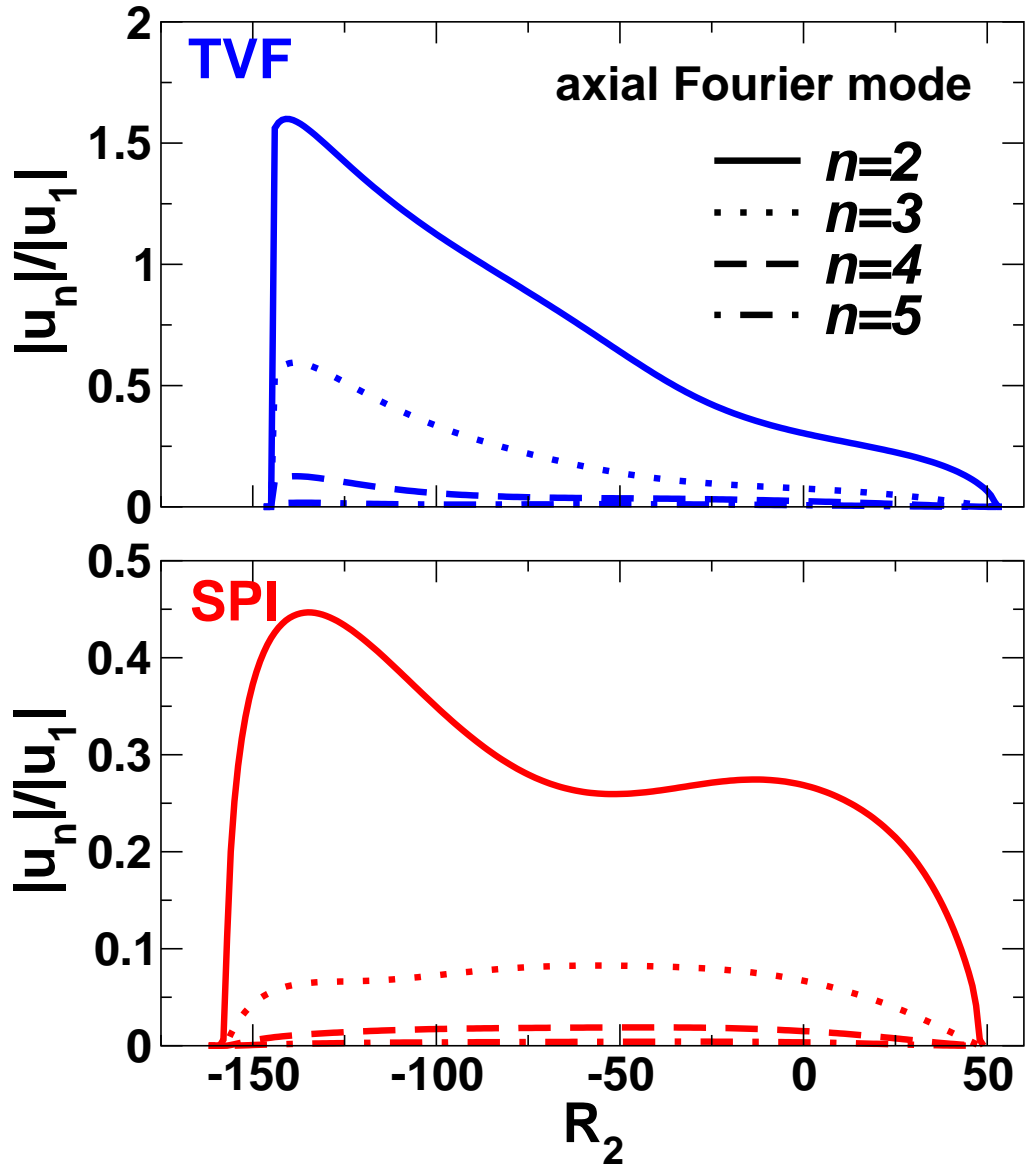}
\end{center}
\begin{quote}
\caption{Anharmonicity of TVF and SPI. The ratios $|u_n/u_1|$ of the axial Fourier 
modes of the profiles of $u(z)$ shown in 
Fig.~\ref{FIG:u-profiles} are displayed here as functions of $R_2$ for fixed 
$R_1=130$. The bifurcation thresholds are located at the zeroes. Parameters are 
$\eta=0.5$, $k=3.927$.}
\label{FIG:anharm}
\end{quote}  
\end{figure}

The variation of the anharmonicity of the vortex profiles can be read off more 
quantitatively from the results of an axial Fourier analysis. To that end we show in
Fig.~\ref{FIG:anharm} the ratios
$|u_n/u_1|$ of the n-th and first axial Fourier modes of the profiles of 
Fig.~\ref{FIG:u-profiles} as a function of $R_2$ for fixed $R_1$. With growing
distances from the bifurcation thresholds at positive and negative $R_2$ the
anharmonicity grows for TVF as well as for SPI. It does so most precipitously 
near the thresholds at negative $R_2$ of about -150 in Fig.~\ref{FIG:anharm}.

At negative $R_2$ the anharmonicity of TVF can be for rapidly counter rotating 
cylinders already close to threshold so large that $|u_2/u_1| > 1$. This 
property reflects the fact that for sufficiently negative $R_2$ Taylor vortices 
are effectively smaller in size than the gap width. There
are two main reasons for this size reduction which are both connected to the
tendency of vortices to have circular shape: ({\it i}) the axial periodicity length 
$\lambda=1.6$ reduces the {\it axial} vortex size relative to the gap and, more
importantly,  
({\it ii}) the TVF intensity is {\it radially} restricted not to extend significantly 
beyond the zero of CCF at $r_0$ since according to the Rayleigh criterion 
$m=0$ radial momentum transport is suppressed by opposite pressure gradients for
$r > r_0$ where the CCF stratification of the squared angular momentum density 
is stable. With $R_2$ becoming more negative $r_0$ moves inwards and the radial 
size of Taylor vortices reduce. 

\begin{figure}[ht]
\begin{center}
\includegraphics[width=14.5cm]{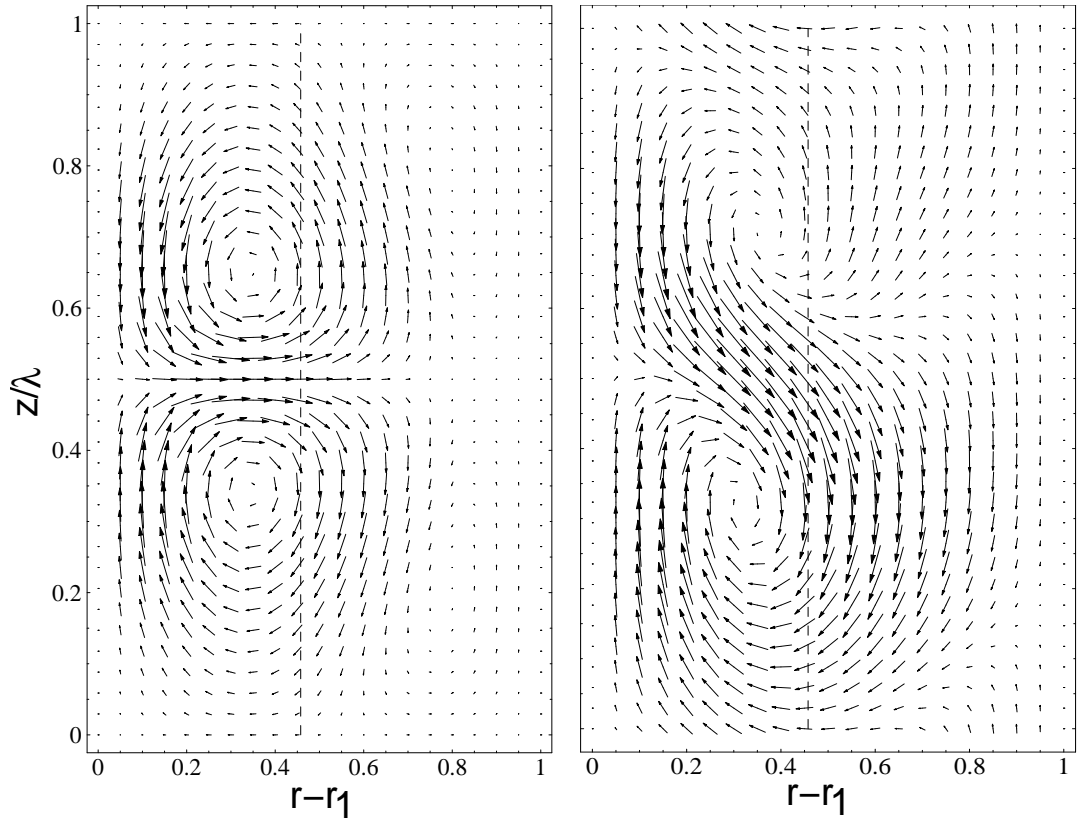}
\end{center}
\caption{Velocity field ($u,w$) of TVF (left) and L-SPI (right) in an 
$r-z$ plane. 
Vertical lines locate the zero of the azimuthal CCF flow $v_{CCF}(r)$.
Parameters are $\eta=0.5$, $k=3.927$, $R_1 = 120$, $R_2 = -100$.}
\label{FIG:arrows_TVF_SPI}
\end{figure}

However, the $m=0$ Rayleigh criterion does not apply to SPI. Their $m \neq0$ radial
momentum transport extends further beyond $r_0$. Therefore SPI vortices fill 
out the whole gap more than Taylor vortices, c.f. Fig.~\ref{FIG:arrows_TVF_SPI}, and 
consequently they are less anharmonic.

\subsubsection{Mirror symmetry breaking of SPI}

TVF shows axial mirror symmetry around the position of maximal
radial outflow,  $z=0.5 \lambda$, in Fig.~\ref{FIG:u-profiles}. In order
to measure the degree to which this symmetry is broken in SPI we have used the
asymmetry parameter
\begin{eqnarray} \label{EQ:P}
P=\frac{\int |u(z')-u(-z')|\,dz'}{\int |u(z')+u(-z')|\,dz'}
\end{eqnarray}
evaluated at mid gap with $z'=0$ locating the largest radial SPI 
outflow at this $r$-value. In this way we found, e.g., for the spirals of 
Fig.~\ref{FIG:u-profiles} that the smallest $P\simeq 0.2$ occurs
for spirals with the smallest frequency $\omega_{min} \simeq 23.4$ at 
$R_2 \simeq -74$. Increasing $R_2$ from this value all the way toward 
the upper SPI bifurcation threshold at $R_2 \simeq 48$ the frequency increases 
but  
$P$ remains roughly unchanged at about 0.2. On the other hand, when decreasing 
$R_2$ from -74 the asymmetry parameter increases with increasing $\omega$ 
up to $P\simeq 1$  close to the lower SPI bifurcation threshold $R_2 \simeq -158$.
Thus, fast propagating spirals at large negative $R_2$ show the largest 
mirror symmetry breaking.

\subsection{Comparison with experimental results}
\label{SEC:CER}

\begin{figure}
\begin{center}
\includegraphics[width=14.5cm]{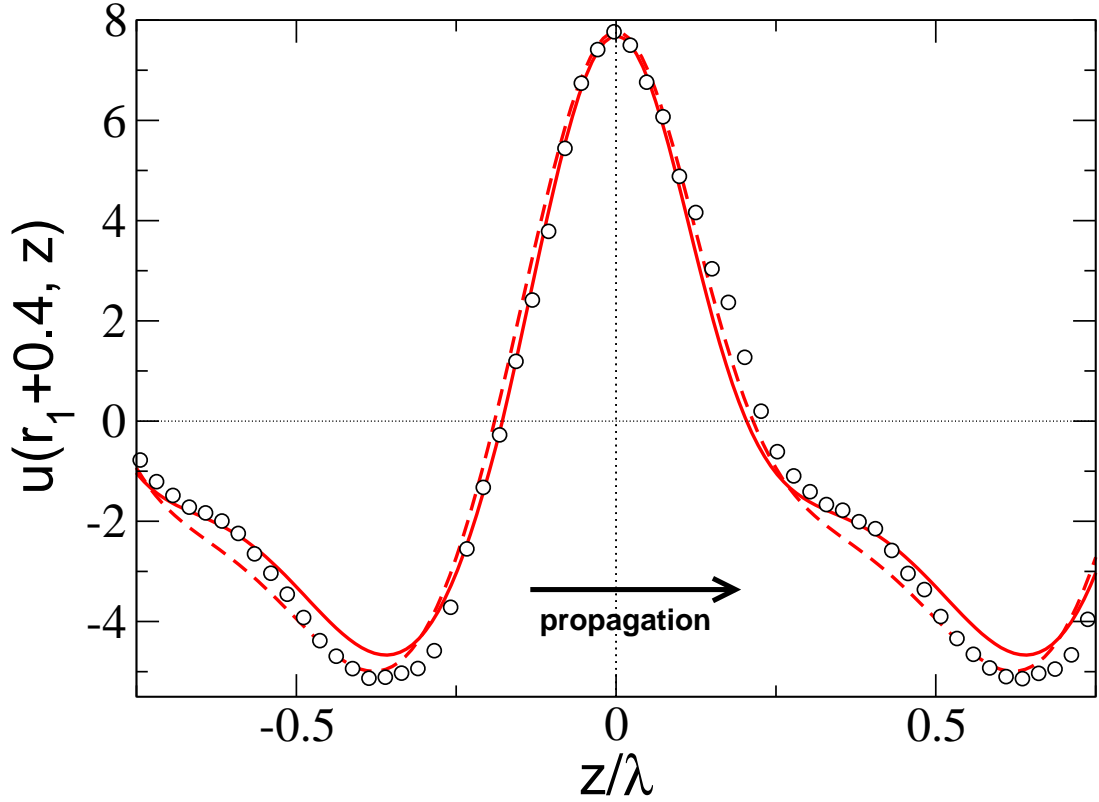}
\end{center}
\begin{quote}
\caption{ 
Comparison of experimental and numerical axial profiles of the radial velocity 
$u(r_1 + 0.4, z)$ of an L-SPI. For better visibility more than one period is shown. 
Symbols and the full line denote Laser-Doppler 
velocimetry measurements \cite{SP99} and numerical simulations, respectively, of a 
Taylor-Couette setup of height $\Gamma=12$ with rigid, non-rotating lids at both 
ends. Both refer to the bulk region at mid height with a common local wavelength of 
$\lambda \simeq 1.76$. There the experimental maximum of $u$ is scaled to our 
simulation 
result. Dashed line refers to a simulation done with axially periodic conditions
imposing a wavelength of $\lambda = 1.6$. Common parameters are
$\eta=0.5$, $R_1=111$ with $R_2=-95$ for the experiments and $R_2=-96$ for the
simulations.}
\label{FIG:exp_num_u}
\end{quote}  
\end{figure}

In order to check our numerical results we made a few comparisons with
experiments. For example, in Fig.~\ref{FIG:exp_num_u} we show the axial profile of the
radial flow $u(z)$ 
of an L-SPI at $r_1 + 0.4$. Symbols denote Laser-Doppler velocimetry 
measurements \cite{SP99} and the full line a numerical simulation, both done in a
setup of height $\Gamma=12$ with rigid, non-rotating lids at both 
ends of the annulus. In each case the spirals were monitored at mid-height of 
the cylinders where they had the common wavelength $\lambda \simeq 1.76$.
Since absolute experimental velocities were not available 
we have scaled the experimental maximum in Fig.~\ref{FIG:exp_num_u} to that of our
simulation (full line). Without knowledge of the experimental error-bars we
consider the agreement between symbols and full line to be satisfactory. 

The dashed line shows a numerical profile obtained for axially periodic 
boundary conditions
imposing the wavelength $\lambda = 1.6$. It differs slightly from the SPI
profile (full line) in the bulk part of the $\Gamma=12$ system with rigid ends.
The difference is presumably related to the fact that the axial flow, and in
particular the mean-flow $w_0$ (\ref{Defw_0}), is different in these two cases
as discussed in Sec.~\ref{SEC:axial-veloc}.

\begin{figure}
\begin{center}
\includegraphics[width=14.5cm]{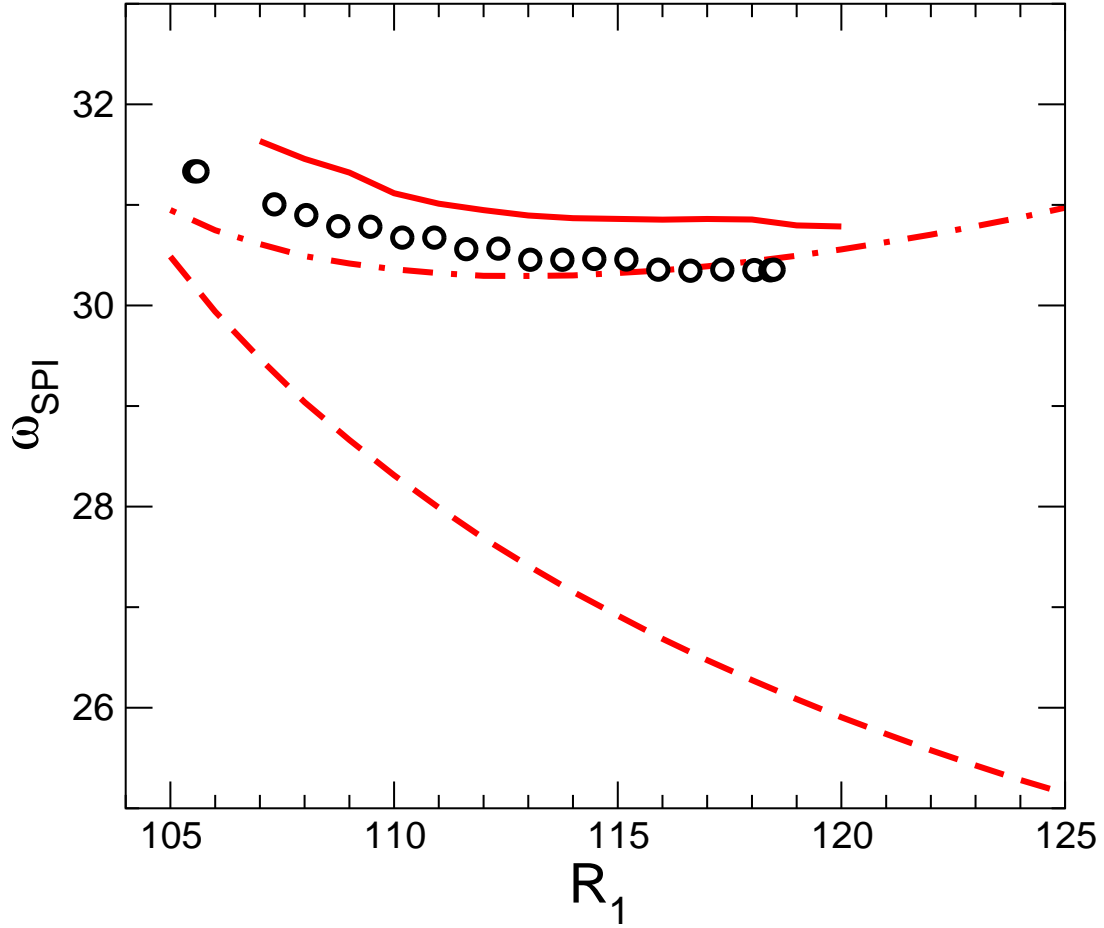}
\end{center}
\begin{quote}
\caption{Comparison of the frequency variation of experimental and numerical 
L-SPI with $R_1$. Symbols and the full line come from Laser-Doppler 
velocimetry measurements \cite{SP99} and numerical simulations, respectively, of a 
Taylor-Couette setup ($\eta$=0.5) of height $\Gamma=12$ with rigid, non-rotating lids 
at both 
ends that enforce the net mean axial flow $\langle w \rangle$ (\ref{Def_w_net}) 
to vanish. Dashed line refers to a simulation done with axially periodic conditions
($\lambda = 1.6$). They allow for a finite Reynolds-stress-sustained 
$\langle w \rangle$ that is negative for our parameters. Upon subtracting this 
Galilean contribution 
$\langle w \rangle k$ from the oscillation frequency under periodic
boundary conditions (dashed line) one obtains the dash-dotted line that lies 
close to the SPI
frequencies with rigid end conditions. Common parameters are $R_2=-96$, however,
$R_2=-100$ for the full line.}
\label{FIG:exp_num_om}
\end{quote}  
\end{figure}

In Fig.~\ref{FIG:exp_num_om} we compare the frequency variation of experimental and 
numerical L-SPI with $R_1$. Symbols and the full line come from Laser-Doppler 
velocimetry measurements \cite{SP99} and numerical simulations, respectively, 
of the aforementioned Taylor-Couette setup ($\eta$=0.5) of height $\Gamma=12$ 
with rigid, non-rotating lids at both ends. Note that not only the frequency values 
of these experimental and numerical SPI states agree reasonably well with each 
other but also
their existence range in $R_1$. Its lower end marks the oscillatory onset.
At the upper end in $R_1$ these SPI lose their stability to TVF --- in 
experiments as well as in the simulations.

However, under axially periodic boundary conditions the existence range of 
stable SPI extends to significantly larger values of $R_1$ lying outside of the
plot range of Fig.~\ref{FIG:exp_num_om}.
The dashed line in Fig.~\ref{FIG:exp_num_om} refers to simulations done with axially 
periodic conditions ($\lambda = 1.6$) that allow for a free propagation of 
phase. And, in addition, they allow the Reynolds-stress-sustained mean 
axial flow $w_0$ (\ref{Defw_0}) to have a finite {\em net} part
$<w>$ (\ref{Def_w_net}) that is negative for our parameters -- c.f. 
Sec.~\ref{SEC:axial-veloc}. In order to compare with the SPI frequencies for 
rigid end conditions we subtract from the oscillation frequencies under 
periodic boundary conditions (dashed line) the pure Galilean contribution 
$\langle w \rangle k$ and obtain the dash-dotted line. Note how close the latter
lies to the SPI frequencies in the system with rigid end conditions. Thus, we
find that the SPI frequency differences \cite{RL93} for the two different end 
boundary conditions are mostly due to whether the Galilean contribution 
$\langle w \rangle k$ is suppressed or not.

\section{External Through-flow}
\label{SEC:ETF}
Here we discuss the influence of an externally imposed axial through-flow on
spiral and on Taylor vortices. Since the effect of an
axial through-flow on TVF has been investigated for $R_2=0$ in several works, 
we focus our investigation on SPI vortices.

The through-flow is enforced by adding in the NSE for the axial velocity 
component a constant 
pressure gradient of size $\partial_z p_{APF}$ throughout the annulus. In the 
absence of any vortex flow, i.e., for sub-critical control parameters
this pressure gradient, $\partial_z p_{APF}$, drives an annular
Poiseuille flow (APF) with a radial profile of the axial through-flow velocity
given by
\begin{equation}\label{wAPF}
w_{APF}(r)= \frac{\partial_z p_{APF}}{4} 
\left[r^2 + 
\frac{1+\eta}{(1-\eta)\ln\eta}\ln r + 
\frac{(1+\eta)\ln(1-\eta)}{(1-\eta)\ln\eta} -
\frac{1}{(1-\eta)^2} \right]
\end{equation}
We checked that our numerical code reproduces this analytical solution
(\ref{wAPF}) of the NSE. We use its mean to define the through-flow Reynolds number 
by
\begin{equation} \label{DefRe}
\left<w_{APF}(r)\right> = Re = 
- \frac{\partial_z p_{APF}}{8} 
\frac{1 - \eta^2 + (1+\eta^2)\ln\eta}{(1-\eta)^2\ln\eta}.
\end{equation}
Hence positive (negative) $Re$ implies an axial flow, $w_{APF}(r)$, in positive 
(negative) $z$-direction. The last equality in Eq.~(\ref{DefRe}) establishes the 
relation between the externally applied additional axial pressure gradient and the
through-flow Reynolds number $Re$. 

\subsection{Counter-rotating cylinders}
Fig.~\ref{FIG:SPIvsRe} shows how the through-flow influences L-SPI, R-SPI, and TVF
at the fixed characteristic driving combination $R_1=120, R_2=-100$ that is
located in Figs.~\ref{FIG:BIF-u} and ~\ref{FIG:BIF-om} in the region C close to the
border to region E. For this parameter combination TVF is unstable when $Re=0$
and it remains unstable in the $Re$-range shown in Fig.~\ref{FIG:SPIvsRe}. This is
of relevance for the through-flow induced transitions between L-SPI and R-SPI (c.f.
further below).

\subsubsection{Bifurcation behavior}
\label{SEC:bif_R_2=-100}

\begin{figure}[ht]
\begin{center}
\includegraphics[width=14.5cm]{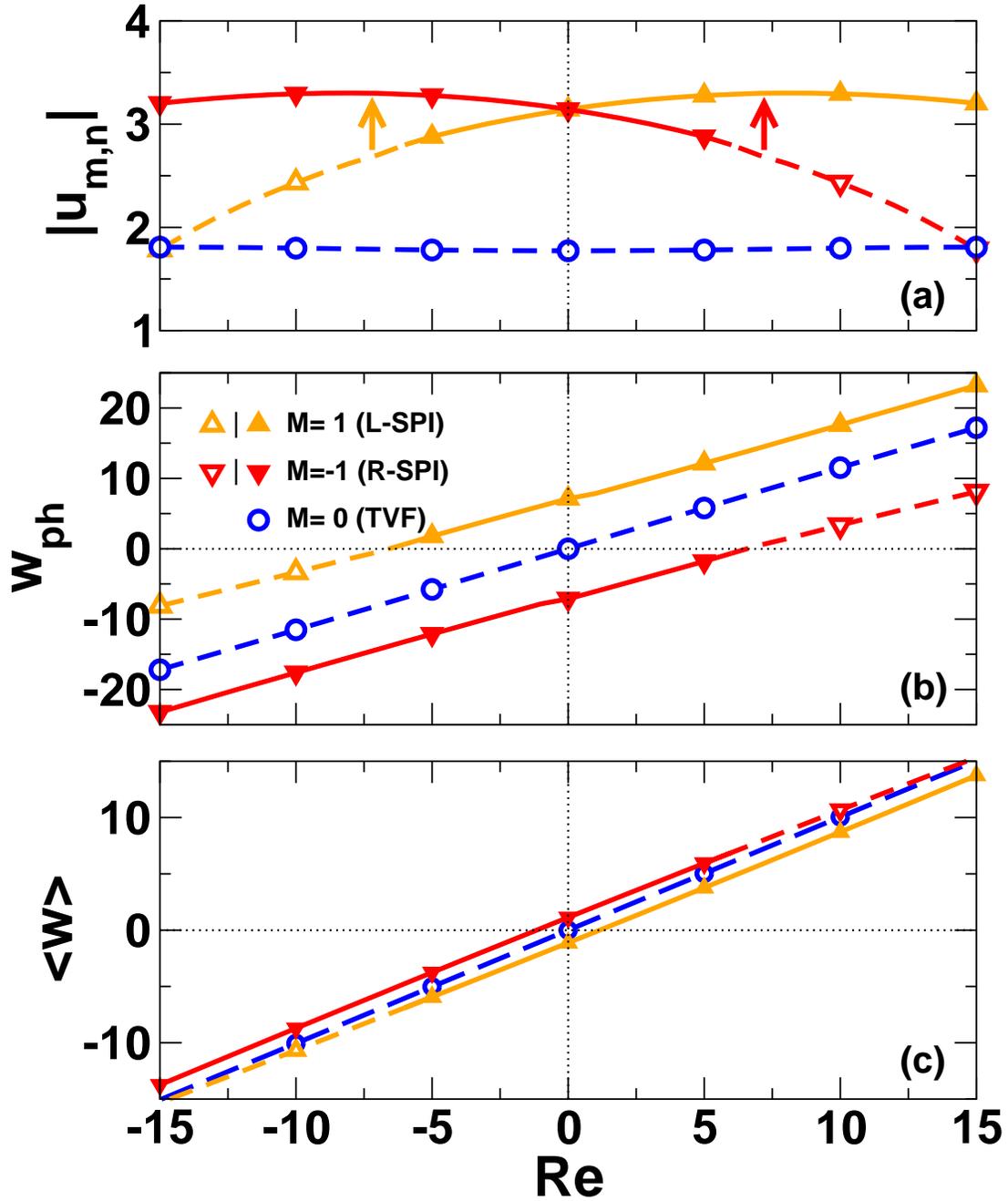}
\end{center}
\begin{quote}
\caption{Influence of an external through-flow on vortex structures. (a) Primary 
Fourier amplitudes of the radial flow field at mid gap for the 
$M=1$ L-SPI ($u_{1,1}$), the $M=-1$ R-SPI ($u_{-1,1}$), and for TVF ($u_{0,1}$).
(b) Axial phase velocity 
$w_{ph}=\omega /k$. (c) Net mean axial flow $\langle w \rangle$ (\ref{Def_w_net}).
Full (dashed) lines with filled (open) symbols refer to stable (unstable) states.  
Arrows indicate transitions after loss of stability, see text for details. TVF is 
unstable in the 
$Re$-range shown here for our parameters $R_1=120, R_2=-100, \eta=0.5, k=3.927$.}
\label{FIG:SPIvsRe}
\end{quote}  
\end{figure}

We present in Fig.~\ref{FIG:SPIvsRe}(a) the 
primary Fourier amplitudes, $|u_{m,n}|$, of the radial flow intensity at mid gap
versus $Re$. These are $|u_{1,1}|$ for the $M=1$ L-SPI, $|u_{-1,1}|$ for the 
$M=-1$ R-SPI, and $|u_{0,1}|$ for TVF. Fig.~\ref{FIG:SPIvsRe}(b) shows their axial 
phase velocity, $w_{ph}=\omega /k$, and Fig.~\ref{FIG:SPIvsRe}(c) shows the 
{\em net} mean axial flow  
\begin{equation} \label{Def_w_net}
\langle w \rangle
 = \frac{1}{\pi (r_2^2-r_1^2)}\int\limits_0^{2\pi}\int\limits_{r_1}^{r_2}
 w(r,\varphi,z,t)rdr\,d\varphi \, .
\end{equation}

For $Re=0$ the two spirals are mirror images of each other: their radial
velocities are the same and all respective axial velocities have the same
magnitude but opposite direction. Note that the SPI Reynolds stresses drive
an axial flow to be discussed further below. Its net mean, $\langle w
\rangle$ (\ref{Def_w_net}), is directed opposite to the phase velocity,
$w_{ph}$, of the respective spiral when $Re=0$. Note, however, the 
difference in size between $|w_{ph}| \simeq 7.1$ and 
$|\langle w \rangle| \simeq 1.1$ \cite{error}.

A finite through-flow breaks the mirror symmetry between the $M=1$ L-SPI and
the $M= -1$ R-SPI. Their radial flow amplitudes evolve with through-flow
as shown in Fig.~\ref{FIG:SPIvsRe}(a). We performed also a 
linear stability analysis of the combined CCF-APF state. It shows that for our
control parameters $R_1=120, R_2=-100$ the amplitudes of the $M=\pm 1$ SPI
solutions go to zero at the bifurcation threshold values of $Re=\mp 19.07$ 
and $Re=\pm 50.95$. The numerical solutions of the full nonlinear NSE showed 
in addition that the L-SPI (R-SPI) is unstable near the first threshold, 
$Re\simeq-19$ ($Re\simeq 19$), and that it is  
stable near the second one, $Re \simeq 50$ ($Re \simeq -50$).

For small through-flow -- say, for $-6 \lesssim  Re \lesssim 6$ in 
Fig.~\ref{FIG:SPIvsRe} -- the two spiral solutions coexist bistably; their particular 
realization depending on initial conditions. However, with increasing $|Re|$ that 
spiral suffers a through-flow enforced loss of stability for which
the phase velocity changes sign. This happens roughly 
when the through-flow has become sufficiently strong to revert an originally 
adverse axial phase propagation. For example, the $M=-1$ R-SPI of 
Fig.~\ref{FIG:SPIvsRe} propagate for small $Re \lesssim 6.6$ axially
downwards (i.e. opposite to the externally imposed through-flow) as for $Re=0$, 
then become stationary, and finally propagate upwards in through-flow direction for 
$Re \gtrsim 6.6$.  
Similarly, by symmetry, the $M=1$ L-SPI propagates in a small negative through-flow
upwards against the through-flow for $Re \gtrsim -6.6$ and downwards, i.e., 
in through-flow direction for $Re \lesssim -6.6$.  
 
The direction of the imposed through-flow is the preferred one for stable phase 
propagation: A spiral that has started at small $|Re|$ to move against the wind
dies out -- or, more precisely, becomes unstable -- when the wind becomes 
sufficiently strong to turn it back. Only that SPI  is stable at large 
$|Re| \gtrsim 7.2$ in Fig.~\ref{FIG:SPIvsRe} that keeps propagating
into the preferred direction of the through-flow. The other one is unstable at 
large $|Re|$.

The through-flow enforced loss of stability of one SPI state
and the transition to the remaining stable one is indicated schematically 
in Fig.~\ref{FIG:SPIvsRe}(a) by vertical arrows. However, we should like to 
stress  that
the transition is somewhat complex extending over the through-flow interval  
$6\lesssim |Re|\lesssim  7.2$ the center of which locates the zero of $w_{ph}$
at $|Re| \simeq  6.6$.
In this interval there are stable, mixed states with
finite L- and R-SPI modes. Their amplitudes seem to vary largely 
continuously with $Re$ (with possibly some saddle-node discontinuity) 
between the pure SPI solutions: the amplitude of the spiral that loses the 
stability competition decreases with growing $|Re|$ towards zero while the amplitude
of the winning one increases from zero to the pure monostable final SPI state. 

Note that since TVF is unstable for the parameters of Fig.~\ref{FIG:SPIvsRe}
it does not offer an alternative transition to a final $M=0$ state as for
the parameters of Sec.~\ref{SEC:inner_cylinder_at_rest}. There, for $R_2=0$,
the through-flow induces a transition to stable TVF rather than to the
stably coexisting SPI with preferred propagation direction. Only when TVF
is eliminated there does the transition occur to the then monostable spiral ---
for details see Sec.~\ref{SEC:inner_cylinder_at_rest}.
  
We also made a few calculations in a regime 
where TVF stably coexists with SPI for counter-rotating cylinders. Also then
the through-flow induces preferably
a transition to stable TVF rather than to the stable SPI state. Thus, when the
through-flow destabilizes, e. g., the $M=-1$ R-SPI then typically the $M=0$ TVF
modes grow rather than the $M=1$ L-SPI modes.

\subsubsection{Axial velocities $w_{ph}, w_0$, and $\langle w \rangle$}
\label{SEC:axial-veloc}
 
In the through-flow range shown in Fig.~\ref{FIG:SPIvsRe} the phase velocity 
$w_{ph}$ and the net mean flow $\langle w \rangle$ vary roughly linearly with
$Re$. The slopes $\partial w_{ph}/\partial Re$ and  
$\partial \langle w \rangle /\partial Re$ for SPI as well as for TVF are 
roughly 1.

While the phase of the $M=\pm 1$ SPI reverts its propagation direction at $Re
\simeq \mp 6.6$ the net mean flow changes sign already at 
$Re \simeq \pm 1.2$. The reversal of the latter does not seem to have any
consequence. But the through-flow enforced reversal of the phase velocity
seems to be responsible for the
destabilization of the SPI that propagate at small $|Re|$ against the wind, i.e.,
into the "wrong" direction.

\begin{figure}[ht]
\begin{center}
\includegraphics[width=14.5cm]{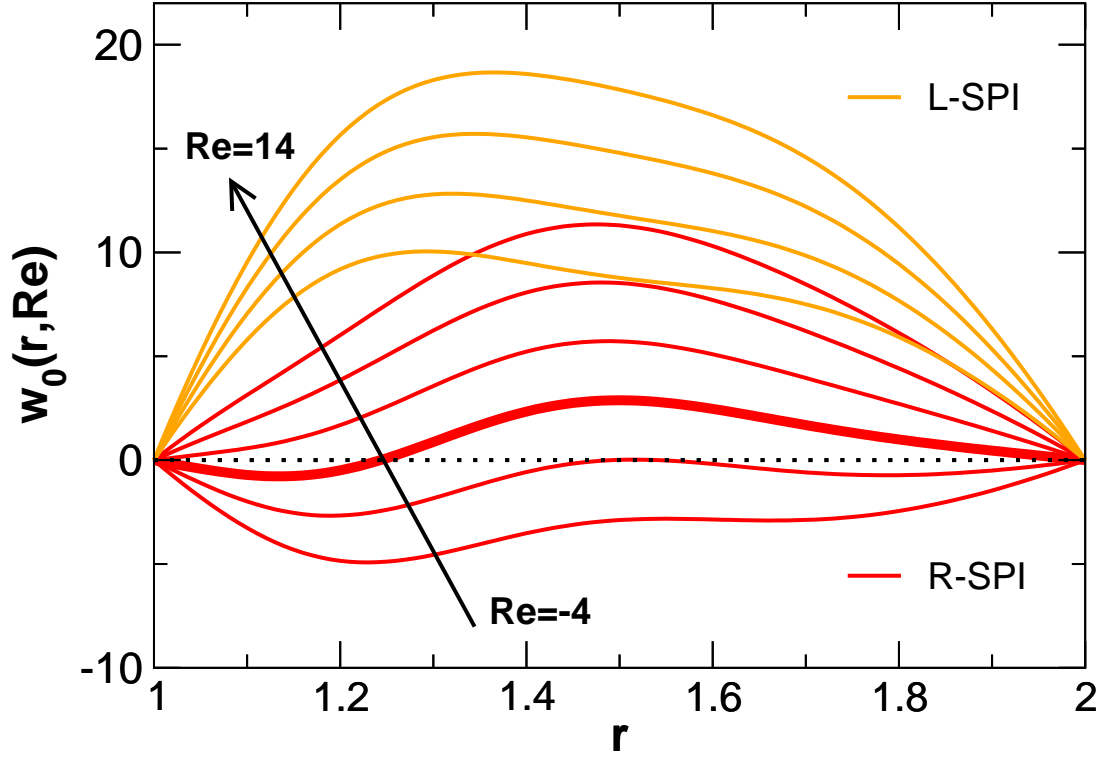}
\end{center}
\begin{quote}
\caption{
Radial profiles of the axial mean flow $w_0(r)$ (\ref{Defw_0})
of spirals shown in Fig.~\ref{FIG:SPIvsRe} for axial
Reynolds numbers $-4\leq Re \leq 14$ increasing in steps of 2. Thick line refers
to $Re=0$. The transition from R- to L-SPI occurs around 
$Re\simeq7$, c.f. text. Parameters are $R_1=120, R_2=-100, \eta=0.5, k=3.927$.}
\label{FIG:w_0}
\end{quote}  
\end{figure}

In Fig.~\ref{FIG:w_0} we show how the radial profiles of the mean axial flow 
\begin{equation} \label{Defw_0}
w_0(r) = \frac{1}{2\pi}\int\limits_0^{2\pi}w(r,\varphi,z,t)\,d\varphi
\end{equation}
of spirals shown in Fig.~\ref{FIG:SPIvsRe} evolve with the through-flow in the
range $-4 \leq Re \leq 14$. We checked that $w_0$ is independent of $z$ and $t$ 
and that our spirals propagating in the externally imposed axial
pressure gradient still have the SPI symmetry, i.e., the flow fields depend on 
$z, \varphi, t$ only via the phase combination $\phi$ (\ref{phase}) with an
oscillation frequency $\omega$ that is modified by the through-flow. Then one
finds from the NSE for the $m=0$ azimuthal mode of the axial velocity field,
\begin{equation} \label{reystr}
  \left(\partial_r + \frac{1}{r}\right) \partial_r  w_0
 =\left(\partial_r + \frac{1}{r}\right) \left( u w \right)_0
 +\partial_z p_0 \, ,
\end{equation}
that the SPI mean flow can be driven by Reynolds stresses and/or by mean axial 
pressure
gradients. For $Re=0$ the pressure is enforced to be axially periodic, hence 
$\partial_z p_0(Re=0) = 0$. So in that case the mean axial flow is driven
solely by the nonlinear Reynolds stresses. They are rather large. For example 
for the R-SPI propagating at 
$Re=0$ in negative $z$-direction with phase velocity $w_{ph} \simeq -7.1$ the maximum
of $w_0(r)$ is about 3, i.e, directed opposite to the phase propagation and
almost half as large in magnitude as $w_{ph}$. The {\em net} mean flow 
$\langle w \rangle$ (\ref{Def_w_net}) is for this case still about 1.1 
and also opposite to $w_{ph}$.

As an aside we mention that rigid axial end
conditions enforce $\langle w \rangle=0$ throughout the annulus. They generate
an adverse axial pressure gradient that compensates the Reynolds stresses 
\cite{ETDS91} so that $w_0$ is practically zero in the bulk part where SPI are
realized. Only in the Ekman region $w_0$ becomes finite showing TVF behavior 
there.

For the R-SPI of Fig.~\ref{FIG:w_0} propagating at $Re>0$ opposite to the
external through-flow the maximal mean flow is located roughly at mid-gap.
However, for the SPI propagating into the direction of the external
through-flow, i.e., the R-SPI for $Re<0$ and the  L-SPI for $Re>0$ the extremum
of $w_0(r)$ is shifted towards the inner cylinder. 
The mean flow profiles of the spirals of Fig.~\ref{FIG:w_0} are given within 
about 5\% by the superposition 
\begin{equation} 
\label{EQ:composition_w_0}
w_0(r;Re) \simeq w_0 (r; Re=0) + w_{APF} (r; Re)
\end{equation}
of the pure, Reynolds stress generated flow $w_0(Re=0)$ of the respective SPI
plus the pure, pressure gradient enforced APF flow $w_{APF}(Re)$
(\ref{wAPF}). This holds for L-SPI as well as for R-SPI, irrespective of whether
they propagate into the direction of the through-flow or against it.

\subsubsection{Spiral profiles}
\begin{figure}[ht]
\begin{center}
\includegraphics[width=14.5cm]{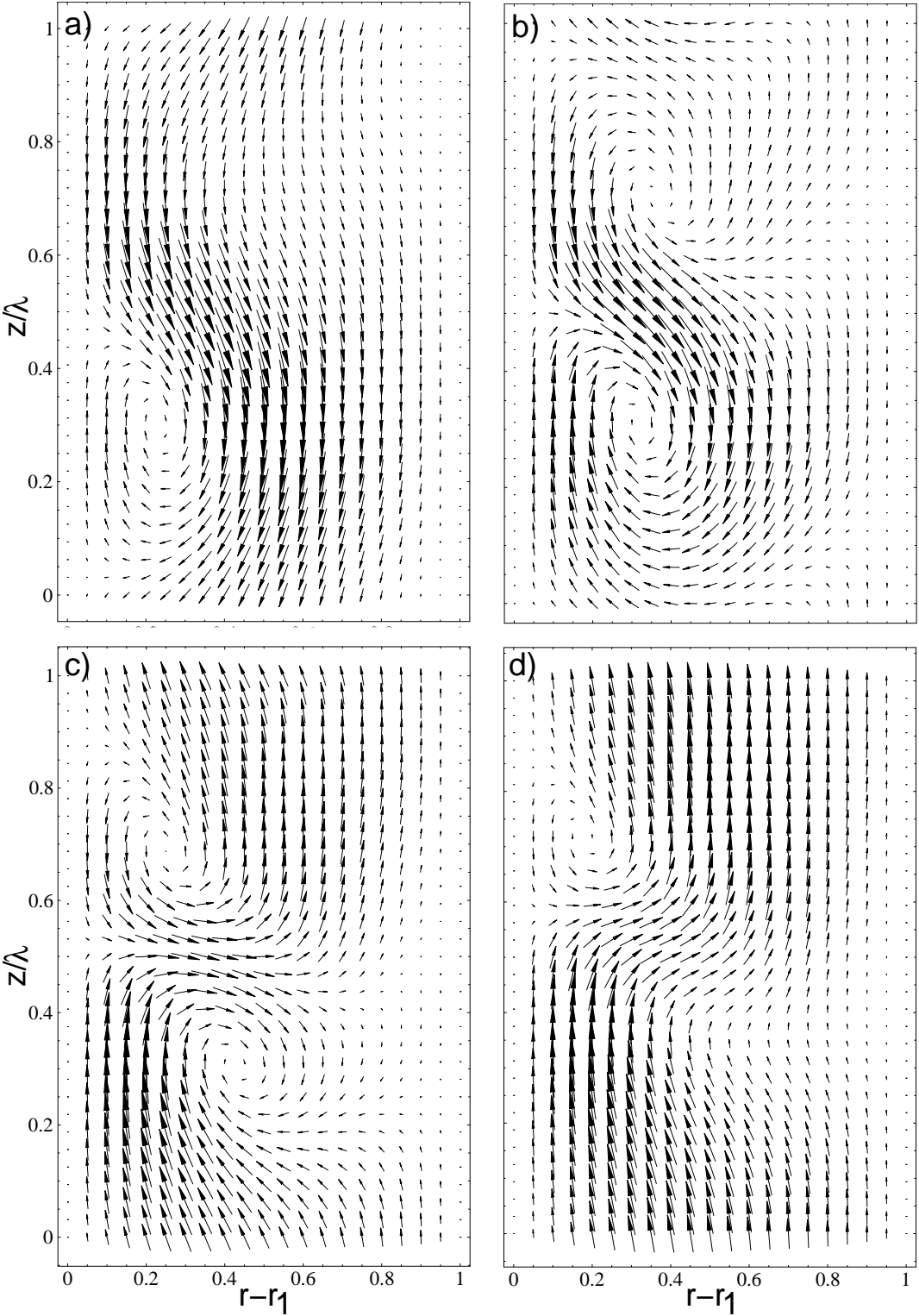}
\end{center}
\begin{quote}
\caption{Velocity field ($u,w$) of L-SPI in an $r-z$ plane for $Re$=-5
(a), 0(b), 5(c), 10(d) . 
Parameters are $\eta=0.5$, $k=3.927$, $R_1 = 120$, $R_2 = -100$.}
\label{FIG:arrows_SPI_Re}
\end{quote}  
\end{figure}

\begin{figure}[ht]
\begin{center}
\includegraphics[width=14.5cm]{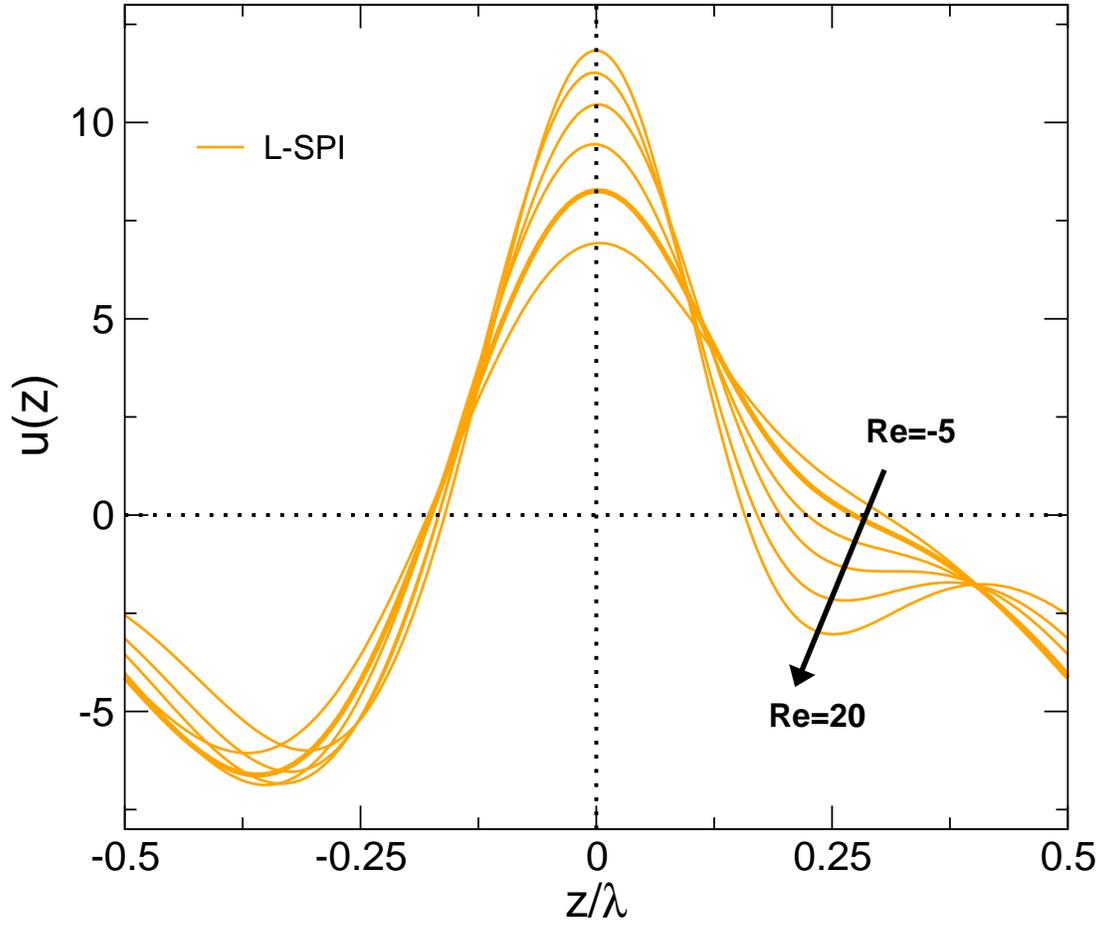}
\end{center}
\begin{quote}
\caption{The effect of an external through-flow on the axial profiles of the 
radial velocity of L-SPI. Lines show $u(z)$ at mid gap position for 
$Re=-5$ to $Re=20$ in steps of 5. Thick one refers to $Re=0$. In each case the 
maximal radial outflow is chosen to lie at $z=0.5 \lambda$. Parameters are 
$R_1 = 130$, $R_2=-100$, $\eta=0.5$, $\lambda=1.6$.}
\label{FIG:u_L-SPI_Re}
\end{quote} 
\end{figure}

The through-flow changes the structure of the SPI. This is documented in 
Figs.~\ref{FIG:arrows_SPI_Re} and \ref{FIG:u_L-SPI_Re}. The arrows in 
Fig.~\ref{FIG:arrows_SPI_Re} representing the $u,w$ vector field of L-SPI in 
the $r-z$ plane show the effect of imposing an axial through-flow that 
increases from $Re=-5$(a) to $Re=10$(d) in steps of 5. Note, however, that the
externally imposed axial pressure gradient does not just add $w_{APF}(r)$ to
the axial velocity field $w$. It also modifies all vector field components
of the SPI. The axial profile of the radial flow $u(z)$ for example is changed
by the through-flow as shown in Fig.~\ref{FIG:u_L-SPI_Re} for increasing $Re$. 
Here the axial asymmetry of the upwards propagating L-SPI is reduced by
steepening up the leading part of $u(z)$ ahead of the wave crests. This
reduction of the mirror-asymmetry of the radial flow of L-SPI grows
somewhat linearly with increasing $Re$. As an aside we mention that on the 
other hand the TVF profiles of 
$u(z)$ become with increasing $Re$ more and more asymmetric --- the mirror
asymmetry parameter $P$ (\ref{EQ:P}) increases for TVF linearly with $Re$.

\subsection{Non-rotating outer cylinder}
\label{SEC:inner_cylinder_at_rest}
We have investigated the influence of an externally imposed axial through-flow on
TVF and SPI also for stationary outer cylinder, $R_2=0$.

\subsubsection{Bifurcation behavior}

\begin{figure}[ht]
\begin{center}
\includegraphics[width=14.5cm]{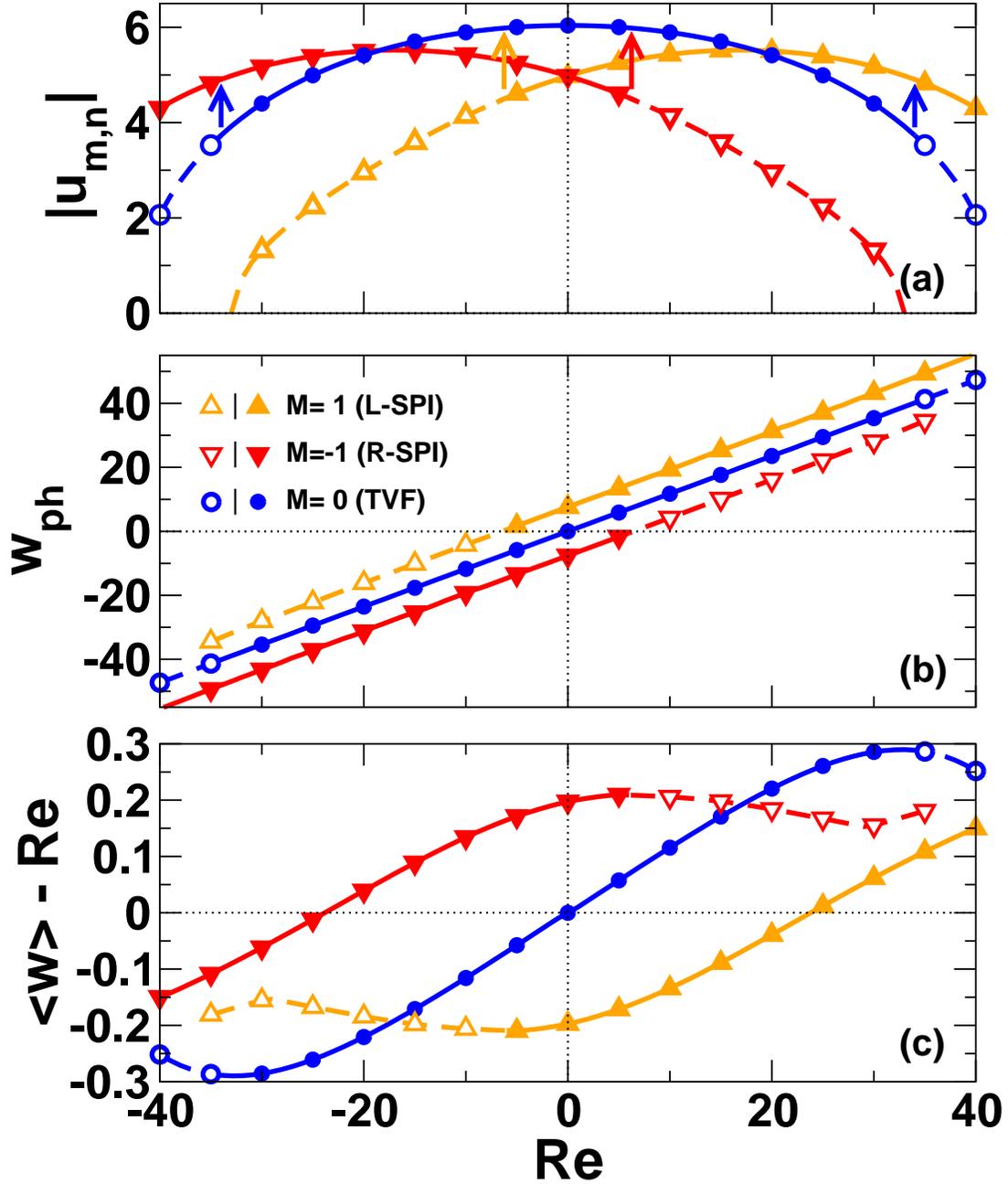}
\end{center}
\begin{quote}
\caption{Influence of an external through-flow on vortex structures. (a) Primary 
Fourier amplitudes of the radial flow field at mid gap for the 
$M=1$ L-SPI ($u_{1,1}$), the $M=-1$ R-SPI ($u_{-1,1}$), and for TVF ($u_{0,1}$).
(b) Axial phase velocity 
$w_{ph}=\omega /k$. (c) Net mean axial flow $\langle w \rangle - Re$. Full 
(dashed) lines with filled (open) symbols refer to stable (unstable) states.  
Arrows indicate transitions after loss of stability, see text for details. 
Parameters are $R_1=100, R_2=0, \eta=0.5, k=3.927$.}
\label{FIG:bif_SPI_TVF_R2=0}
\end{quote}  
\end{figure}

In Fig.~\ref{FIG:bif_SPI_TVF_R2=0} we show the bifurcation behavior of TVF
and SPI as a function of through-flow Reynolds number $Re$ for $R_2=0$,
$R_1=100$.  This parameter combination lies well within the region E of
Fig.~\ref{FIG:BIF-u} in which TVF, L-SPI, and R-SPI are all stable at
$Re=0$.
 
Switching on the through-flow one sees in Fig.~\ref{FIG:bif_SPI_TVF_R2=0}(a)
how the dominant modes of these vortex structures vary with $Re$. That SPI
loses its stability for which the through-flow enforces a reversal of the
phase propagation as in the case of counter rotating cylinders
(Fig.~\ref{FIG:SPIvsRe}). Thus, also here the direction of the imposed
through-flow is the preferred one for stable SPI at large $|Re|$. A spiral
that has started at small $|Re|$ to move against the through-flow becomes
unstable when the latter becomes sufficiently strong to turn it back. On the
other hand that SPI remains stable at large $|Re|$ that keeps propagating
into the preferred direction of the through-flow.
 
As in Fig.~\ref{FIG:SPIvsRe} the loss of stability takes place in the vicinity 
of the Reynolds number where the
axial phase velocity $w_{ph}$ [Fig.~\ref{FIG:bif_SPI_TVF_R2=0}(b)] of the
respective SPI goes through zero. This happens in
Fig.~\ref{FIG:bif_SPI_TVF_R2=0} for the $M=\pm 1$ SPI at $Re \simeq \mp
6.4$. However here we found the 
transition from the then unstable SPI to occur to the stable TVF solution 
[c.f. arrows in Fig.~\ref{FIG:bif_SPI_TVF_R2=0}(a)] rather than to the other
stable SPI.

We have also investigated briefly the situation where the TVF solution was
numerically eliminated (here, suppressing $m=0$ modes of the
$u$-field at mid-gap position turned out to be an efficient way to globally
reduce TVF towards zero).
Also then the SPI that is unfavored by the through-flow loses its
stability. However, with TVF being unavailable as final state the
transition occurs in this case to the favored SPI in a way that seems to be
similar to the one described in Sec.~\ref{SEC:bif_R_2=-100}.

Without the above described numerically imposed mode restriction TVF is stable
for moderate through-flow rates while at sufficiently large $|Re|$ SPI 
are stable \cite{BP90,LDM92,TS94b}. For our parameters TVF decays at 
$Re \simeq \pm34$ into a $M=\pm 1$ SPI as indicated by arrows in
Fig.~\ref{FIG:bif_SPI_TVF_R2=0}(a).

For small through-flow the phase velocity 
$w_{ph}$ and the net mean flow $\langle w \rangle$ vary roughly linearly with
$Re$. The initial slopes $\partial w_{ph}/\partial Re$ and  
$\partial \langle w \rangle /\partial Re$ are for SPI as well as for TVF  
roughly 1. However, at larger $Re$ one sees in 
Fig.~\ref{FIG:bif_SPI_TVF_R2=0}(c) that in particular $\langle w \rangle$ shows
nonlinear corrections. 

\subsubsection{Phase diagram}

\begin{figure}[ht]
\begin{center}
\includegraphics[width=14.5cm]{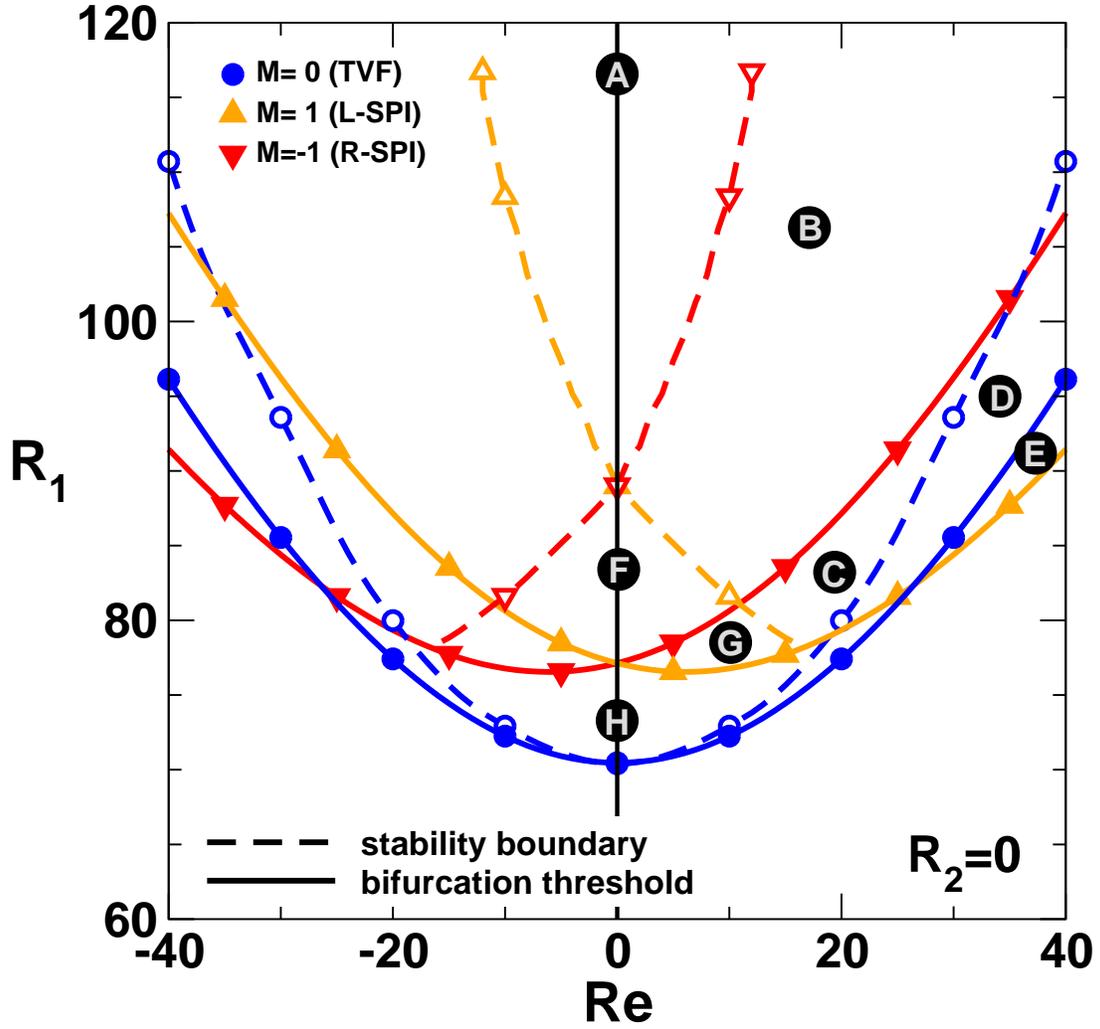}
\end{center}
\begin{quote}
\caption{$R_1 - Re$ phase diagram of TVF, R-SPI, and L-SPI for stationary 
outer cylinder. Solid lines represent linear stability thresholds of the basic
state, i.e., bifurcation thresholds of the respective vortex solutions out of
the combined CCF-APF. Dashed lines are stability boundaries of the vortex
states. The phase diagram is symmetric under $Re \to -Re$. Parameters are 
$R_2=0, \eta=0.5, k=3.927$.}
\begin{tabular}{|c|c|c|c|c|c|c|c|c|}
\hline
region & A & B & C & D & E & F & G & H \\\hline\hline
TVF    & s & s & s & u & - & s & s & s \\\hline
R-SPI  & s & u & - & - & - & u & - & - \\\hline
L-SPI  & s & s & s & s & s & u & u & - \\\hline
\end{tabular}
s: stable; u: unstable; -: nonexistent.
\label{FIG:phasediagram}
\end{quote}  
\end{figure}

Fig.~\ref{FIG:phasediagram} shows the phase diagram of TVF, R-SPI, and L-SPI
for stationary outer cylinder in
the control parameter plane spanned by $Re$ and $R_1$. The {\em existence} range
of the vortex states is bounded from below by the 
bifurcation threshold (full line in Fig.~\ref{FIG:phasediagram}) of the 
respective vortex solution out of the combined CCF-APF basic state. These
bifurcation thresholds result from a linear stability analysis of the CCF-APF 
state \cite{PLH03}. The one for TVF increases quadratically for small $Re$.
Also the SPI threshold curves in Fig.~\ref{FIG:phasediagram} have a somewhat 
parabolic shape, however with minima shifted to finite $Re$. Thus, the threshold
for L-SPI first decreases for small positive $Re$ but eventually increases at
larger $Re$. By symmetry the R-SPI threshold curve in 
Fig.~\ref{FIG:phasediagram} is the mirror image under $Re \to -Re$ of the L-SPI 
threshold curve. Hence small through-flow destabilizes (stabilizes) the CCF-APF
state against spirals that propagate into (against) the through-flow direction.

Note that for small $Re$ in Fig.~\ref{FIG:phasediagram} TVF bifurcates first
when increasing $R_1$. But for sufficiently large $Re$ the bifurcation sequence
of TVF and SPI is reversed since the bifurcation threshold for TVF curves up 
faster with increasing $Re$ than the one for L-SPI. After their intersection 
stable SPI bifurcate first out of the CCF-APF state. Hence, for example in 
region E of Fig.~\ref{FIG:phasediagram} only stable L-SPI exist, in region D
TVF exists but only as unstable solution, and in region B they exist bistably.
 
The dashed lines in Fig.~\ref{FIG:phasediagram} are stability boundaries of the
vortex solutions. Different regions of Fig.~\ref{FIG:phasediagram} between
various stability boundaries and bifurcation thresholds are identified with 
the respective stability properties of the vortex states in the caption of 
Fig.~\ref{FIG:phasediagram}.

\section{Summary}
We have numerically simulated vortex flow structures of different azimuthal wave
numbers $M$ in the Taylor-Couette system with
counter-rotating as well as with co-rotating cylinders. In particular we have
investigated the effect of an externally imposed axial through-flow
on the spatio-temporal properties and on the bifurcation behavior of 
$M=1$ L-spirals, $M=-1$ R-spirals, and $M=0$ Taylor vortices. 

To that end we first have determined for zero
through-flow, $Re=0$, the bifurcation surfaces of the appropriate order 
parameters characterizing SPI and TVF solutions over the $R_1 - R_2$ control 
parameter plane of the inner and outer cylinder's Reynolds numbers. 
For the parameter combinations explored in this work these
bifurcations out of the basic CCF state are forward and their order of
appearance determines the stability of the respective bifurcating vortex state: 
the vortex solution that bifurcates second 
is unstable. But it eventually becomes stable with 
increasing distance from the bifurcation threshold so that, e.g., for larger 
$R_1$ there is a large region in the $R_1 - R_2$ plane with bistability of TVF 
and SPI. In particular the existence region of stable SPI extends for axially 
periodic boundary conditions even to positive $R_2$ with co-rotating cylinders.
Unstable solution branches were obtained by selectively suppressing 
destabilizing modes. Stable ribbons, i.e., nonlinear combinations of $M=\pm 1$ 
spirals were not found.

Simulations of axially finite systems with rigid, 
non-rotating lids showed in good agreement with experiments how the 
stable existence range of SPI is reduced by stationary Ekman vortices 
which suppress phase propagation at the two ends. Also the frequencies and the
wave profiles of the spiral vortices in the
bulk of the numerical and experimental systems agreed well with each other.
Spiral profiles obtained for periodic and rigid end conditions do not differ 
much. On the other hand, the respective frequencies differ basically by the 
Galilean contribution $\langle w \rangle k$. Here $\langle w \rangle $ is the net
axial mean flow that the nonlinear Reynolds stresses of a spiral with axial wave 
number $k$ sustains with
axially periodic end conditions but not with impermeable ends. 

Furthermore, we showed how the phenomenon of rigid body rotation of spirals can 
be understood quantitatively in terms of the passive advection dynamics of  
$M = \pm 1$ vortex perturbations whose lines of constant phase are oriented 
obliquely to the azimuthal CCF. The onset spiral frequency is
the mean rotation rate of the CCF, albeit weighted appropriately by the 
critical eigenfunctions with the consequence that L-SPI as well as R-SPI rotate
into the same direction as the inner cylinder. The nonlinear SPI frequencies are 
typically smaller than the linear ones but do not deviate substantially from them.
    
A finite through-flow breaks the mirror symmetry between the L-SPI and the R-SPI
and it changes the structure of the SPI. The externally imposed axial 
pressure gradient does not just add the annular Poiseuille flow $w_{APF}(r)$ to
the axial velocity field. It  modifies the SPI structure, e.g., the profiles of 
the radial flow in a characteristic way. 

For $Re=0$ L-SPI propagate axially upwards and R-SPI downwards.
When they are initially stable they continue to coexist bistably
for small through-flow. However, they are no longer mirror images of each other
and their phase velocities differ by an amount $\propto Re$. Then, with 
increasing $|Re|$ that spiral loses its stability for which 
the through-flow enforces the phase velocity to change direction. Only that SPI
is stable at large $|Re|$ that keeps propagating
into the preferred direction of the through-flow. The other one is unstable at 
large $|Re|$.

The SPI that loses stability upon reverting its propagation direction --- i.e. 
the R-SPI (L-SPI) for positive (negative) $Re$ --- preferentially undergoes a 
transition to propagating TVF provided the latter is available as {\em stable} 
vortex state. Otherwise the transition is to the then monostable L-SPI (R-SPI).
Such a situation was explored in detail for negative $R_2$ where TVF was
unstable and for other parameter combinations where the TVF solution was
eliminated numerically. 

Also the situation where initially at $Re=0$ all three vortex solutions are 
stable was elucidated for different $R_1 -R_2$ parameter combinations and in 
more detail for stationary outer cylinder, $R_2=0$. Here, a complete phase 
diagram was determined in the control parameter plane spanned by $Re$ and $R_1$.
We found that small through-flow destabilizes (stabilizes) the basic CCF-APF
state against spirals that propagate into (against) the through-flow direction.
For sufficiently large $Re$ the bifurcation sequence
of TVF and SPI is reversed since the bifurcation threshold for TVF curves up 
faster with increasing $Re$ than the one for L-SPI. After their intersection 
stable SPI bifurcate first out of the CCF-APF state. Then there opens up a
region at sufficiently large positive $Re$ in which only stable L-SPI but no
Taylor vortices exist for stationary outer cylinder.

\section*{Acknowledgments}   
We thank A. Schulz for communicating the
experimental data referred to in this paper.

\clearpage



\begin{thebibliography}{99}
\bibitem{error}
In Fig. 5(a) of Ref. \cite{HL} the quantity labelled "Axial mean flow 
$-\langle w \rangle$" is too small by a factor of $2\pi$.
\end{thebibliography}
\end{document}